# Space-time Trends in U.S. Meteorological Droughts

**Authors and Affiliations:**

Poulomi Ganguli[1] and Auroop R. Ganguly[1]*

[1]Sustainability and Data Sciences Laboratory, Civil and Environmental Engineering, Northeastern University, 360 Huntington Avenue, Boston, MA, 02115, USA

* Corresponding author:

Sustainability and Data Sciences Laboratory, Civil and Environmental Engineering, Northeastern University, 360 Huntington Avenue, Boston, MA, 02115, USA

Email: poulomizca@gmail.com; Phone: +1(617)-373-3395

**Abstract:**

Understanding droughts in a climate context remains a major challenge. Over the United States, different choices of observations and metrics have often produced diametrically opposite insights. This paper focuses on understanding and characterizing meteorological droughts from station measurements of precipitation. The Standardized Precipitation Index is computed and analyzed to obtain drought severity, duration and frequency. Average drought severity trends are found to be uncertain and data-dependent. Furthermore, the mean and spatial variance do not show any discernible non-stationary behavior. However, the spatial coverage of extreme meteorological droughts in the United States exhibits an increasing trend over nearly all of the last century. Furthermore, the coverage over the last half decade exceeds that of the dust bowl era. Previous literature suggests that climate extremes do not necessarily follow the trends or uncertainties exhibited by the averages. While this possibility has been suggested for droughts, this paper for the first time clearly delineates and differentiates the trends in the mean, variability and extremes of meteorological droughts in the United States, and uncovers the trends in the spatial coverage of extremes. Multiple data sets, as well as years exhibiting large, and possibly anomalous, droughts are carefully examined to characterize trends and uncertainties. Nonlinear dependence among meteorological drought attributes necessitates the use of copula-based tools from probability theory. Severity-duration-frequency curves are generated to demonstrate how these insights may be translated to design and policy.

## 1. Introduction

Drought is a recurrent problem in multiple regions of the CONterminous United States (CONUS). Heat waves and droughts have caused damages of around $210.1 billion dollars during 1980-2011 in the US, and have ranked second highest after tropical cyclones in terms of financial losses (Smith and Katz, 2013). The 2012-2013 drought affected approximately two-third of the CONUS, and caused $40 billion financial losses in the agricultural sector and reduced national hydropower generation by 10% (DHS, 2015, pp. 1-87). Climate change has been linked to increasing severity and duration, while influencing spatiotemporal variability, of droughts in the coming decades (Burke et al., 2006; Cook et al., 2015; Dai, 2013).

### 1.1 Conflicting and Uncertain Insights

Uncertainties in drought characterization inhibit the translation of scientific understanding into resilience policy (Trenberth et al., 2014; Tsakiris et al., 2007; Van Loon, 2015; Vogel et al., 2015). According to the Fifth Assessment Report (AR5) of the Intergovernmental Panel on Climate Change (IPCC Change, 2013), our confidence in characterizing, projecting and attributing droughts has steadily decreased from 2007 (IPCC Fourth Assessment Report, or AR4) and 2012 (IPCC SREX, or Special Report on Extremes) to 2013 (AR5). Arguably among the most important climate-related hydrologic insight, highlighted in successive IPCC reports including the latest AR5 (IPCC Change, 2013), is that climate change causes dry regions to get drier and wet regions to get wetter. However, even this supposedly confident claim has been disputed by recent findings (Greve et al., 2014). Two recent papers (Dai, 2013; Sheffield et al., 2012) provided diametrically opposite insights about the sign of the trend in droughts globally and in the U.S. over the last half a century. A subsequent paper (Trenberth et al., 2014) suggested that the different insights in the two back-to-back papers were caused by the use of different data and metrics.

The hydrologic community has noted a slight decrease in observed mean U.S. drought trends and considerable uncertainty in the assessments (Orlowsky and Seneviratne, 2013; Sheffield et al., 2012; Trenberth et al., 2014; Van Loon, 2015; Vogel et al., 2015). However, extreme drought events have been occurring, including over the last several decades, causing widespread misery. A recent article on droughts (Diffenbaugh et al., 2015) mentions: "changes in the atmospheric mean state may not be reflective of changes in the risk of extreme events (including atmospheric configurations conducive to precipitation extremes)". This paper examines the hypothesis that extreme droughts in the U.S. have been exhibiting a statistically discernible increasing trend, even though mean drought trends may be uncertain and may even show a slightly decreasing trend.

*1.2 Droughts over the Continental United States*

Over the CONUS, no evidence was found of increasing frequency, spatial extent or severity of droughts until the 1990s (Karl and Heim, 1990). Recent studies suggest identical insights (Table S1), although a few regional exceptions exist (Andreadis and Lettenmaier, 2006; Easterling et al., 2007; Seager et al., 2012, 2007). The dichotomy between mean versus extreme drought severity (or intensity) is apparent given the fact that the existing literature shows a decrease in average drought severity over the CONUS in aggregate (Table S1), but increasing trends in extreme drought severity regionally (Cole et al., 2002; Woodhouse and Overpeck, 1998). Space–time drought characterizations over the US (Andreadis et al., 2005; Mishra and Singh, 2010; Rajsekhar et al., 2014; Sheffield et al., 2009) have tended to focus on the areal extent and trends in mean drought state however lesser attentions have been paid to changes in variance and drought persistence. The recent California drought is claimed to be exceptionally severe in the past millennium (Griffin and Anchukaitis, 2014) with no rain recorded in downtown San Francisco in January for the first time since 1893 (Los Angeles Times, 2015). A recent study (Diffenbaugh et al., 2015) showed that although there has not been a substantial change in the probability of either negative or moderately negative precipitation anomalies over California in recent decades, the occurrence of drought years has been greater in the past two decades than in the preceding century. A recent National Climatic Data Center (NCDC) report (NCDC, 2012) asserts that over the CONUS, the 2012 drought is comparable to the 1930s dust-bowl drought. As discussed previously, while it may seem counterintuitive, it is not unusual for the hydro-meteorological events to show differential patterns in mean versus extremes since changes in the atmospheric mean state do not necessarily reflect changes in the risk of extreme events (Diffenbaugh et al., 2015). However, the existing literature on US droughts does not attempt to comprehensively distinguish between trends in mean drought severity versus its extremes, although similar analyses have been performed for other hydrologic or climatic extremes (Elsner et al., 2014, 2008; Emanuel, 2005; O'Gorman, 2014).

*1.3 Choice of a Drought Index*

Recent research has examined droughts, especially in a climate context, through the Palmer Drought Severity Index (PDSI) and variants (Table S1). The PDSI estimates relative changes in soil moisture using a physical water balance approximation (Palmer, 1968, 1965). However, the PDSI is subject to estimation and generalization challenges. In addition, any tendency to treat this index as an independent indicator of drought may lead to false associations and physical insights.

The challenges in PDSI estimation from data are illustrated through the previously mentioned divergent insights about drought trends (Dai, 2013; Sheffield et al., 2012; Trenberth et al., 2014). Another approach of drought management at a catchment scale is the use of hydrological models, such as, the Variable Infiltration Capacity model, VIC (Andreadis and Lettenmaier, 2006). However, low flows are often poorly simulated by commonly used hydrological models (Smakhtin, 2001; Staudinger et al., 2011). Furthermore, complex parameterizations of land-atmosphere and subsurface processes introduce considerable uncertainties in the model simulations (Li et al., 2007). In-situ measurements of soil moisture may be thought to provide point estimates of past PDSI but such measurements are available only with inadequate spatial and temporal coverage (Robock et al., 2000). The website for PDSI of the University Center for Atmospheric Research (UCAR) [https://climatedataguide.ucar.edu/climate-data/palmer-drought-severity-index-pdsi] mentions two limitations. According to the website, the PDSI is (a) "not as comparable across regions as the Standardized Precipitation Index (SPI)", although "this can be alleviated by using the self-calibrating PDSI" and (b) the PDSI "lacks multi-timescale features of indices like the SPI, making it difficult to correlate with specific water resources like runoff, snowpack, reservoir storage, etc.". Finally, since the PDSI is a measure of relative changes in soil moisture based on water balance approximations, any attempt to treat this index as an independent variable, or relate it to other indicators may lead to misleading physical insights. The Standardized Precipitation Index (SPI) is perhaps the most widely operationally used drought monitoring and assessment index in the U.S. and many other regions across the globe. Indices based exclusively on precipitation, including the SPI, do not consider the complexity of land surface processes and cannot directly account for the impacts of evaporation or transpiration on soil moisture. This may be a particularly serious drawback under warming conditions, or other changes in regional hydrometeorology. However, among the advantages are the direct and exclusive relation to a measured variable, precipitation, despite the large uncertainties in precipitation measurements and hence among precipitation datasets (Fekete et al., 2004; Trenberth et al., 2014). In addition, compared to the PDSI, the SPI is less complex, provides early warning and is able to simulate multi-time scale aspects of droughts (Hayes, 2006; Mo, 2011). Based on these considerations, this paper examines droughts through the SPI index. Since the SPI considers precipitation alone, the paper examines "meteorological droughts" exclusively. However, agricultural or hydrological droughts may be indirectly related to meteorological droughts, especially since the last may be a precursor to the other two.

*1.4 State of the Art on Translation to Actionable Insights*

Critical challenges remain not only in drought characterization, but also in the translation of drought attributes to information relevant for monitoring, attribution, early warning, resources planning and infrastructures resilience. Differences in the trends of the mean versus extremes makes the situation even more complex. In addition, the variability in drought attributes, as well as any non-stationary behavior (Hughes et al., 2012; Kundzewicz, 2011; Verdon-Kidd and Kiem, 2010) therein, is potentially more important for planning than average trends (Milly et al., 2008). One objective way to inform drought planning is to construct severity-duration-frequency (SDF) curves (*e.g.*, as developed for Texas (Rajsekhar et al., 2015, 2014) similar to the now common intensity-duration-frequency (IDF) curves for precipitation extremes (Aron et al., 1987; Huff et al., 1992; Yarnell, 1935). However, just as for IDF, interdependence among drought attributes cannot be ignored for appropriate translation to risk management (Salvadori et al., 2013) through SDF curves. One reason for the lack of wider adoption of SDF curves may be the difficulty in translating joint risks given the nonlinear interdependence among multiple drought attributes. Recent literature recommends the applicability of copula-based models for improving multivariate drought characterization (Mishra and Singh, 2011). In fact, SDF curves have been used in different parts of the world to determine the relationship between drought hazard and vulnerability (Todisco et al., 2013). In the US, only a few studies reported construction of SDF curves but the analyses were primarily limited to watershed (Bellamy et al., 2012; Wang et al., 2011) or regional scale drought assessments (Rajsekhar et al., 2015, 2014). (Andreadis et al., 2005) developed severity-area-duration curves for the CONUS from 1920 – 2003 based on VIC simulated soil moisture percentiles.

*1.5 Research Questions and Hypotheses*

Given the aforementioned challenges in drought characterization and translation, a focused study is needed to comprehensively analyze drought trends and spatiotemporal patterns, with a particular need to delineate mean versus extreme trends.

This paper examines the following *primary hypothesis* over the U.S. (i.e., the CONUS):

*Droughts and their attributes such as severity, duration, frequency and spatial coverage over the U.S. have not been exhibiting any significant upward/downward trend.*

A thorough examination of the broad hypothesis above leads to the following research questions, especially in light of the existing (and often conflicting) insights in the literature:

- o  Do mean and extreme drought trends differ over the CONUS and in regions thereof?

- o  How do extreme drought events influence the overall trend?

- Are meteorological drought patterns consistent across observational datasets?

- Do regional drought patterns persist in regions that are currently drought–prone, and do drought patterns emerge in new areas?

- Is the variance in drought attributes (mean and extreme) stationary and could this influence the examination of the broad hypothesis above?

- What is the relation between drought attributes (severity, duration, frequency and spatial coverage) and how do interdependencies between these attributes impact design curves?

To address these questions, we analyze the space-time variability of meteorological droughts, defined in terms of precipitation deficits over the multiple regions in the CONUS. Investigating meteorological droughts is important by itself, and also because they act as precursors to longer lasting and often more consequential agricultural and hydrological droughts (Haslinger et al., 2014; Mo, 2011; Wilhite et al., 2014). As a case in point, in a recent study based on a hydro-meteorological time series of 44 Austrian catchments, (Van Loon and Laaha, 2015) claimed that hydrological drought properties are largely controlled by average catchment wetness, which is in turn represented by mean annual precipitation. We design our analysis to account for the uncertainty arising from choice of datasets as well as the influence of outlying drought years on drought trends. We examine trends based on precipitation datasets using rigorous statistical evaluations but move beyond analyzing "anomalies" of precipitation (Diffenbaugh et al., 2015; Easterling et al., 2007). While our characterizations of average drought severity are based on relatively straightforward statistical approaches that consider space time patterns and trends, characterization of extreme drought severity is conceptually similar to the precipitation extremes considering "$T$-year return period" (*i.e.,* severity expected to occur once every $T$-years) of the event. However, interdependence between drought severity and duration may be nonlinear, which need to be taken into account for characterizing extremes. Thus, we have used a copula-based approach to characterize extreme drought severity. The use of copulas has been suggested for precisely this purpose in the literature  (Nelsen, 2006, Sklar, 1959) and there are numerous examples of copula applications in the context of drought management (Hao et al., 2014; Kao and Govindaraju, 2010; Maity et al., 2013; Wong et al., 2009). Further, as a step toward translation to actionable insights, copula-based drought severity-duration-frequency curves are generated, which in turn explicitly consider interdependence among drought attributes.

Our insights on the attributes of meteorological droughts can inform drought mitigation (Wilhite and Svoboda, 2000). However, the insights are conditioned on the quality of station-level precipitation data that are used in the analyses. The remainder of the paper is organized as follows: Section 2 discusses data and methods, sections 3 and 4 present results and discussions respectively, section 5 concludes our paper.

## 2. Data and Methods

### 2.1. Study Area and Data

The observed precipitation datasets have been carefully selected to (a) ensure coverage over the CONUS throughout most of the last century, (b) delineate the trends and patterns that are data-dependent versus those that are agnostic to the choice of a dataset (and hence potentially more general), and (c) quantify the influence of extremely large drought years, which may be considered as "outlying" cases.

Two different ground-based observational precipitation datasets are selected for the continental US (20°N–50°N; 125°W–60°W).

- The first dataset is a subset of the United States Historical Climatology Network (USHCN Version 2.5) instrumental monthly precipitation records (Menne et al., 2010) at 1218 meteorological stations from 1926–2013. USHCN Version 2.5 represents one of the best available datasets for investigating long-term changes, as the stations are chosen based on the length and quality of the data (Karl et al., 1986; Vose et al., 2003). Since the credibility of a meteorological drought index largely depends on the quality of observed precipitation data, we select only those stations that have less than six months of missing records. Following this criterion, the total number of usable stations in the USHCN dataset is reduced to 616, of which missing values constitute less than 6% of the record during the analysis period of 1926-2013. The missing values in a particular month are imputed using time series interpolation, which is one of the commonly used methods to estimate missing records in hydrology (Mizumura, 1985; Price et al., 2000). A shape-preserving piecewise cubic polynomial function (Fritsch and Carlson, 1980; Hyman, 1983) was chosen for the interpolation. Unlike other interpolants (such as, linear and spline), this interpolation function is able to preserve local monotonic trends in the dataset such that extreme artifacts are not introduced. In places where interpolation of precipitation fields occasionally produces negative values, missing values are replaced by the median monthly precipitation values for that year.

- The second dataset is the hourly precipitation record from the NCDC's archives (NCDC DSI-3240) for approximately 5,500 stations from 1950–2009. We selected 1023 meteorological stations, including only those that have less than 10% missing data in any given year during the analysis period of 1950–2009. The hourly precipitation records are aggregated to monthly totals for the selected stations. The NCDC dataset contains a larger number of observations than the USHCN data. The NCDC dataset excludes what may be viewed as the "outlying" past drought years, specifically the dust bowl of the 1930s. To enable an appropriate comparison, we do not consider the recent (2012s to now) extreme droughts either from the NCDC data. We are therefore left with one dataset without the extreme droughts in the beginning and at the end of the period of coverage and another dataset where these extreme years are included. Insights from the two datasets may help us delineate the trends that agree across datasets, as well as ascertain the influence of the extreme drought years. Thus, we construct a three-way comparison of the two datasets as follows:

I.  USHCN version 2.5 (from 1926–2013) with "outlying drought years" (hereafter referred to as *"D1-out"*) – which include drought events from 1930-31, 1934, 1936 and 1939-40 (Karl and Heim*, 1990*) as well as the ongoing 2012's California droughts.

II. Dataset 1 (from 1950–2009) without the outlying drought years (hereafter referred to as *"D1-No-out"*) as described above.

III. NCDC DSI 3240 (from 1950–2009), which is already without the outlying drought years (hereafter referred to as *D2-No-out*).

The D1-out dataset is split into two equal 44-years' time slices: 1926–1969 and 1970–2013. The choice of comparison periods was based on the prior literature which suggests most of the anthropogenic warming has occurred since the 1970s (Change, 2013; Dai et al., 2004; Peterson et al., 2008). For the other two datasets ("D1-No-out" and "D2-No-out"), we consider an even 30-year split: 1950–1979 versus 1980–2009. First, we perform a head-to-head aggregate comparison across the three datasets to be able to assess the ability of insights to generalize across observational datasets and characterize the extent to which the presence of extreme (and what may be considered outlying) drought years influence the insights. Next, we analyze the dataset which has the longest temporal coverage, specifically D1-out, in detail to extract detailed insights. The analysis is performed over nine NCDC regions, which are mutually exclusive and when taken together sub–divide the continental U.S. into nine "climatologically homogeneous regions" (Karl and Koss, 1984; Karl and Koscielny, 1982), as well as the continental U.S. as a whole. These nine regions are the ones commonly used in the previous literature (Easterling et al., 2007; Soulé and Yin, 1995) and were originally delineated based on principal component analysis of gridded PDSI data. Figure S1 shows the spatial distributions of the meteorological stations corresponding to the USHCN and the NCDC datasets over the nine climatic regions.

*2.2 Methods*

2.2.1 STANDARDIZED PRECIPITATION INDEX (SPI) IN METEOROLOGICAL DROUGHT DETECTION

As discussed previously, this paper focuses on meteorological droughts, which in turn may be defined as a gradual accumulation of precipitation deficit (Svoboda et al. 2012) and modeled using the Standardized Precipitation Index (SPI) (Guttman, 1999; McKee et al., 1993). Compared to other drought index, SPI has the advantage of flexibility to measure precipitation deficit at multiple time scales. SPI represents the number of standard deviations (following a statistical distribution transformed to a normal distribution) above or below that an event happens to be from the long run mean (Sims et al, 2002). To estimate SPI, at a "$n$–month" time scale (hence, SPI–$n$), an accumulation window of $n$-months is applied to a given monthly precipitation time series, following which a statistical distribution is fitted. Although McKee et al. (1993) originally used a Gamma distribution function, other distribution functions could also be used to fit the data. (Stagge et al., 2015) compared a suite of candidate probability distributions at the continental scale focusing on Europe and found that the two-parameter Gamma distribution is suitable for general use when calculating SPI across all accumulation periods and regions. In this paper, based on previous literature on US droughts (Logan et al., 2010; Mo, 2011; Mo and Schemm, 2008), we used the two-parameter Gamma distribution to fit precipitation time series aggregated over $n=6$ months. According to (Svoboda et al., 2012), SPI at 6-months (SPI-6) is appropriate to analyze seasonal to medium trends in precipitation.

## 2.2.2 DROUGHT EVENT AND ASSOCIATED DROUGHT PROPERTY IDENTIFICATION

Drought properties are derived using threshold methods based on the statistical theory of runs *Yevjevich* (1983) for analyzing sequential time series. Baseline (average) precipitation conditions are represented by SPI = 0; negative SPI values denote drier than normal conditions, and positive SPI values indicate wetter than average conditions. A drought event is identified when an uninterrupted sequence of SPI values (at monthly time scales) remains equal to or below the 20[th] percentile of the SPI distribution over the period analyzed at a specific site (Svoboda, 2002).

A single drought class provides information about monthly drought conditions but not about drought duration. We characterize four attributes of each drought event:

(*i*) Drought duration (*D*): Number of consecutive months when SPI remains equal to or below the 20[th] percentile threshold (McKee et al. 1993).

(*ii*) Deficit volume or severity (*S*): Cumulative values of SPI within a drought event (McKee et al., 1993), *i.e*,

$$S_i = -\sum_{t=1}^{D} SPI_{\{i,t\}} \quad i = 1, \ldots, n \qquad (1)$$

where $t = 1$ starts with the first month of the drought event and continues till end of that event (over the duration, $D$) for $i$=6-month (for SPI-6, as explained in section 2.2.1) time scales. Drought severity has the units of equivalent month (McKee et al., 1993).

(*iii*) Drought Persistence: Since drought is an event-based phenomenon, it may extend for more than a season and even years. While understanding seasonal drought persistence is important for short-term water resources planning, such as in agriculture and energy sectors (Ford and Labosier, 2014), decadal and multi-decadal drought persistence has implications for long-term water management, such as designs of hydraulic infrastructure (Borgomeo et al., 2014). Hence, we consider drought persistence with (Ford and Labosier, 2014) and without (Mo and Schemm, 2008; Sheffield et al., 2009) seasonal influence. When computed without accounting for seasonal influence, a drought event is considered as persistent if it lasts at least a year (12 months) or longer. Based on this criteria, we define persistence probability of droughts as the total number of drought events with duration of at least one year or more, divided by the total number of drought events during the analysis time frame. For example, if a station has two drought events with duration greater than or equal to 1-year, out of a total of 18 drought events, the persistence probability of drought is 2/18, or 0.11. Seasonal persistence is defined as the number of events exhibiting drought persistence across seasons at each station. Subsequently, seasonal persistence probability is calculated as the seasonal persistence divides by the total number of events that occurred during the first season at a particular station (Ford and Labosier, 2014). Seasonal persistence is assessed for summer-to- fall (June - November) and winter -to-spring (December - May) droughts.

(iv) Spatial extent: The percentage of stations considered to be experiencing drought if the SPI value for a particular month and location reaches below the specified threshold, *i.e.*:

$$A = \sum_{i=1}^{N_{Station}} \mathbf{1}\left\{Z_{i,t} \leq Z_{thr}\right\}.A_i \bigg/ \sum_{i=1}^{N_{Station}} A_i \tag{2}$$

where $\mathbf{1}\{\psi\}$ is a logical indicator function of set $\psi \in \{0 \text{ if false and 1 if true}\}$ and $\psi = f\left(Z, Z_{thr}, A\right)$, $Z_{i,t}$ is the SPI value at month $t$ for a station, $A_i$ is the influence area of the station $i$, $Z_{thr}$ is the threshold value of SPI for drought identification for the particular location.

Figure S2 gives an example of the identification of these drought attributes using a meteorological station (location: longitude -117.08° and latitude 32.64°) in California. The drought event (for example event 1) starts at month $t_i$, when the SPI value drops below the threshold limit, has a deficit volume or severity $S_i$ (run-sum) that lasts over the deficit duration, $D$ (run-length) and ends on month $t_e$.

The nature of spatial variability of drought persistence is investigated considering three scenarios: growth, emergence and (decreasing) trend. A persistent drought is considered to have "emerged" when no such droughts are noted during the first half (say, 1926–1969) but at least one or more occur during the second half (e.g., 1970–2013). Growth of drought events is defined as the presence of at least one persistent drought during the first half (1926–1969) and an increase in the count of the number of drought events in the current time window (1970–2013) compared to the prior window of equal size. Drought persistence is considered to be decreasing when the number of such droughts decreased in the second period compared to the first. The relative frequency of persistent drought between the two time periods is compared using a paired $t$-test at $\alpha_L = 0.10$ significance level (Hogg and Tanis, 1977).

While analyses are performed on individual station levels, a simplified aggregation is performed in the ArcGIS 10.2 platform for exploratory visual analysis. To understand spatial variations well, the point estimates of meteorological observations are smoothed using ordinary kriging at a horizontal grid resolution of 0.5° with the commonly used spherical semi-variogram model (Ahmed et al., 2014). Kriging has been widely applied in the literature for spatial analysis of droughts (Alamgir et al., 2015; Kim et al., 2002; Wang et al., 2014). Since the overall results are based on trends in individual station-based observations, we must emphasize that the spatial smoothing does not influence our analysis or insights.

Average drought attributes at individual station level are characterized based on simple statistical methods based on the prior literature (Table S1), while the characterization of extreme drought attributes needs to be more involved owing the interdependence among attributes. This paper uses copula-based approaches to characterize extreme droughts and for translation to SDF curves.

## 2.2.3  TRANSLATION TO DROUGHT SEVERITY–DURATION–FREQUENCY (SDF) CURVES WITH COPULAS

The joint dependence of drought properties, severity and duration, are modeled using the copula function, which in turn enables the quantification of a functional relationship between the $n$-dimensional distribution function and its univariate marginal cumulative distributions  (Nelsen, 2006). Copulas are selected as the tools of choice owing to their ability to characterize complete dependence structures (in this case, among drought attributes) irrespective of the nature of the marginal distributions (Candela et al., 2014; Genest and Favre, 2007; Halwatura et al., 2014; Hao et al., 2014; Nelsen, 2006). Copula-based approaches are also used in this paper for the development of SDF curves which can be the bases for decisions and planning (Dalezios et al., 2000; Halwatura et al., 2014; Shiau and Modarres, 2009).

The marginal distributions of drought properties are modeled using more than ten drought events. Based on the prior literature (Fleig et al., 2006; Rajsekhar et al., 2014; Shiau and Modarres, 2009) a suite of statistical distributions such as Lognormal, Gamma and Weibull may be fitted to marginal distributions of drought severity. The duration time series tends to be discrete and multiple events of the same period may re-occur within the analysis time span, which may result in statistical ties (Serinaldi et al., 2009). Here we fit either the Exponential distribution or a Lognormal distribution. We check the performance of the marginal distribution fits through distance based statistics, specifically, the Akaike Information Criteria (AIC), between theoretical and rank-based empirical distributions. Validity of the marginal distribution fits are checked via classic bootstrap-based ($n = 1000$ replications) Kolmogorov-Smirnov (K-S) goodness–of–fit test ($\alpha = 0.05$ significance level) (Zucchini, 2000).

For modeling joint distributions, we employ four different families of copulas previously used in hydrology (Genest and Favre, 2007; Kao and Govindaraju, 2010; Solari and Losada, 2011): Frank, Gumbel-Hougaard, Plackett, and Student's *t*. The parameters of the copula function are estimated using the Maximum Pseudo-likelihood (MPL) method (Kojadinovic and Yan, 2010). We test goodness-of-fit of the copula models using Cramér-von Mises distance (*i.e.*, the integrated squared difference) between empirical and parametric copula distributions and statistical $p$-values obtained via parametric bootstrap (at $n = 250$ Bootstrap replications and $\alpha = 0.05$ significance level) approach (Berg, 2009; Genest et al., 2009).

Drought Severity-Duration-Frequency (SDF) relations are derived with copula-based conditional return periods (Janga Reddy and Ganguli, 2012; Shiau and Modarres, 2009). Figure S3 shows the flowchart for generating copula-based SDF relations from observed precipitation data. The analytical formulae for the conditional distribution form of the Student's $t$ and the Gumbel-Hougaard copula families are available in the literature (Cherubini et al., 2004; Joe, 1997). The conditional distributions of the other two copula families are derived using first order partial differentiation of the copula distribution with respect to the conditioning variable (Joe, 1997).

The hydrologic insights, whether for average or extreme attributes of droughts, need to be examined through appropriate local and field significance tests. In subsequent subsections we describe statistical significance tests for trends in extreme drought properties.

## 2.2.4 DETECTION OF TRENDS IN THE SEVERITY OF EXTREMES OVER NCDC CLIMATE REGIONS

Drought severity values for the two time segments (1926–1969 vs. 1970–2013 and 1950–1979 vs. 1980–2009) are obtained for return periods $T = 10$-year and $T = 100$-year conditional on drought duration, $d = 1, 2 \dots 12$ months. We consider only those common stations (count = 179 in USHCN version 2.5, and count = 312 in DSI-3240) that contain more than ten drought events during both halves of the periods. To investigate the existence of nonstationarity in drought severity time series, a *local* significance test [following (Livezey and Chen, 1983)] at individual stations is performed for the properties of the first two moments, mean and variance, over the two time windows. Significant differences in median and slope changes are also reported. We examine significant differences in the median at coincident stations using the nonparametric Wilcoxon sign-rank test within a paired dataset at $\alpha_L = 0.05$ significance level. Differences in slope between two time periods are evaluated using the Student's *t* test (at $\alpha_L = 0.05$) for each station. The differences in significance of variance are evaluated using *F*-test (at $\alpha_L = 0.05$) for homogeneity of variance.

## 2.2.5 TESTING FOR FIELD SIGNIFICANCE

Hydrologic and climatological data are expected to exhibit considerable spatial correlation and consequently the results of *local* significant tests cannot be assumed to be independent (Livezey and Chen, 1983; Wilks, 2006). First, since precipitation time series often display strong seasonality, we compute spatial cross-correlations using the standardized monthly anomaly (deviations of monthly values from individual monthly means over the study period divided by corresponding standard deviations: Figure S4) of precipitation time series. The spatial patterns of cross correlation are estimated using Kendall's $\tau$, which measures rank correlations. Figure S5 indicates that in general regional precipitation time series is positively cross-correlated. Next, collective statistical significance (or, field significance) is evaluated using a false discovery rate (FDR)-based approach (Benjamini and Hochberg, 1995; Benjamini and Yekutieli, 2001), which has been compared to other methods and has been shown to be a powerful test and relatively insensitive to spatial interdependence among sites (Khaliq et al., 2009; Ventura et al., 2004). Field significance tests are performed in this paper at the same significance levels as their locally identified trends. While we use $\alpha = 0.10$ significance level to test the trends in persistence probability, we use $\alpha = 0.05$ to test the trends in rest of the statistics (*i.e.*, mean, median and slope). At $\alpha = 0.05$, less than 1% of stations show trends in persistence probability, so we relax the significance level to $\alpha = 0.10$ to increase the power of the test. The results are presented in detail in section 3, discussed with possible mechanistic explanations in section 4 and summarized in section 5.

## 3. Results

### 3.1 Generalizable Trends, Sensitivity to Data Choice, and Presence of Outlying Years

Figure 1 (*top panel, middle and right*) presents a three-way comparison of the differences (recent minus past climatology) in annual average precipitation and average severity among the three datasets, D1-Out, D1-No-out and D2-No-out as described in Section 2.1, D1 and D2 are the two original observational datasets, while the suffixes Out and No-out denote datasets with and without outlying years respectively. A quantitative test of the similarity between recent and past climatology is evaluated using the Wilcoxon rank-sum test. Stipples denote stations where the differences in annual average precipitation are locally significant ($\alpha_L = 0.05$). Figure 1 (*top panel, left*) shows increases in annual average precipitation in D1-Out dataset over most of the regions (Table S2). Locally significant increases in mean precipitation are observed in 34% of the stations, with some larger and spatially coherent increases, especially over parts of the Midwestern, eastern Southcentral and Northeast regions. Decreases in precipitation vary across regions. Only 2% of the stations show a significant decrease, particularly over Florida and Southern California. In contrast, the exclusion of outlying years significantly changes the nature of insights. Changes in annual average precipitation in D1-No-out (Figure 1; *top panel, middle*) shows wetter condition throughout however this is less conspicuous than that of D1-Out, especially over the Midwest and Northeast regions. Upward trends are field significant in most of the regions except in Central, Southeast and West (Table S2) United States. Locally significant (but not field significant) decreasing trends in mean precipitation are observed over the Southeast and Northwest regions. On the other hand, D2-No-out shows (statistically) significant decreases in mean precipitation over more than 17% of stations. The pattern is field significant over the West, Northwest, East-north Central, Southeast, and Northeast regions. Further, field significant increase in mean precipitation is noted over the South, Southwest and West-north Central regions.

The trends in average drought severity (weighted by the duration of each event) in D1-Out show drying patterns over the Southeast, Southwest, Northwest, Northern plains, parts of the Midwest and Appalachia (region in the eastern US that stretches from west of the Catskill Mountains of New York to northern Alabama, Mississippi and Georgia) regions. A spatially coherent wetting pattern is observed over Southcentral (part of Texas), the Great Lakes, and Northeast regions. In D1-No-out and D2-No-out, an increase in average severity is observed over the Southwest and parts of the Southeast. However, none of these trends are found to be field significant. Overall, both dry and wet patterns are intensified over the shorter period (1950–2009), suggesting that the inclusion of outlying years in general counterbalances the intensity of change. The spatial pattern of D2-No-out shows intense drying and wetting patches overall. In general, there is a tendency towards drying trends with less spatial correspondence in the D1 dataset (*i.e.*, both D1-Out and D1-No-out). The disagreement in sign is prominent over Northeast, Northwest, Southwest and West regions. The most notable exceptions occur over the Northeast, where decreases in severity are field significant (Table S2). Taken together the following broader insights emerge regarding mean climatology: (*i*) trends in annual mean precipitation and average meteorological drought severity are more sensitive to the choice of dataset rather than inclusion/exclusion of outlying years, and (*ii*) including outlying years reduces the intensity of changes in average drought severity.

Next, we examine the trends in extreme drought severity. We obtain drought severity values at each time segments from copula-based conditional return periods, $T = 10$ years and $T = 100$-years at durations $d = 1, 2 \ldots, 12$ months. Details of the copula-based conditional return periods and development of severity-duration-frequency curves are described in the Method section. Based on goodness-of-fit tests, drought severity values are found to be best modeled by the Lognormal distribution for the majority of the stations, followed by the Weibull, and then the Gamma distributions. Drought durations are found to be best modeled by the Exponential and the Lognormal distributions. Table S3 shows the results of the marginal distribution fits of a few randomly selected stations with varying number of drought events in four climatic regions for different datasets. The analysis indicates satisfactory fits between the empirical (observed) and theoretical (modeled) distributions. For modeling joint distributions, the Student's *t* copula emerged as the best copula family to handle joint distributions between drought properties for most of the stations in all datasets. For D1-Out and D1-No-out datasets Student's *t* copula performs well for all stations. For the D2-No-out dataset, the contributions from the other copula families are only 3% (1% for each of the families). For Student's *t* copula, multiple values are checked for the degrees of freedom, $\vartheta = 2, 3, \ldots, 10$ (Mashal and Zeevi, 2002). Table S4 shows the results of copula fits for select stations from Table S3. The expression for the conditional return period for Student's *t* copula is analytically solvable at $\vartheta = 6$ (for datasets D1-Out and D1-No-out) and 8 (for dataset D2-No-out) respectively, while satisfying goodness-of-fit test.

Figures 2 and 3 show changes in the spatial coverage of central tendency (median) and the slope of drought severity for extremes (characterized by 10 and 100-year return period) respectively. An analysis of the trends in the median severity for D1-Out suggests increasing trends in the spatial extent of extremes over most of the regions, especially in the West (Figure 2; Table S5 and S6). However, both D1-No-out and D2-No-out show a field significant downward trend across the overall CONUS region. A few regional differences exist (Table S5 and S6) such as, for changes in average severity, a field significant upward trend is observed in the Southwest in D1-No Out dataset at 10- and 100-year return periods. Similarly, the Northeast at 10-year return period and the Southwest at 100-year return period exhibit significant upward trends. Likewise, in the D2-No Out dataset, Central, East-north Central, Northeast, South, Southwest and West-north Central regions show field significant upward trends at all (10- and 100-year) return periods. For changes in slopes, Central, East-north Central, Northeast, Southeast, South, and Southwest regions show field significant upward trends in D1-No Out dataset. On the other hand, in D2-No Out dataset, all regions except West show field significant upward trends in slope of severity at 10-year return period. Further, at 100-year return period, all regions exhibit field significant upward trends. However in general, the number of stations with significant downward trends exceeds the number of stations with (significant) upward trends (Table S5 and Table S6). Our results corroborate the existing literature in terms of decreasing trends in drought severity overall (Andreadis and Lettenmaier, 2006; Easterling et al., 2007).

The changes in the slope of severity (Figure 3) are found to be sensitive to both the choice of the dataset and the inclusion or exclusion of outlying years. The trends in the slope of severity in D1-Out show a field significant upward trend in the spatial coverage of extremes over most of the regions (Table S5 and S6). Disagreement in the nature of the trend is prominent over Central and Northeast regions between D1-No-out and D2-No-out datasets. However, on the whole, Figure 3 suggests robust increases in the slopes of extremes over drought prone areas such as the South, Southwest and Southeast, which are further influenced by the inclusion of outlying years (Table S5 and S6).

The results of trends in the median severity of extremes suggest at least two key insights: (*i*) when the outlying years are included, extreme droughts tend to show more spatial coverage. This implies that recent droughts have more spatial coverage for extremes than the droughts in the past (*i.e.,* the 1930s and 50s'). (*ii*) The impact of outlying years generally dominates over the choice of data set. Further, trends in the slopes of extreme drought severity suggest (*i*) an upward trend in slopes in recent years, which is in turn amplified by the inclusion of outlying years, and (*ii*) upon excluding outlying years, the field significant upward trend is mostly contained within drought-prone parts of the country, specifically the South, Southeast and Southwest regions.

*3.2 No Significant Changes in Drought Persistence*

Multiple regions over North America exhibit persistent droughts extending over seasons and even across years. Understanding the persistence of droughts is important for sectors ranging from agriculture (Basso and Ritchie, 2014; Lobell et al., 2014) to energy (Hightower and Pierce, 2008; Palmer and Lund, 1986). The likelihoods of drought persistence between the two time periods (1970–2013 versus 1926–1969) are compared using spatial plots of persistence probability (Figure 4). The black filled circles indicate locally significant ($\alpha_L = 0.10$) differences in the persistence probability of drought in two 44-year periods. The spatial pattern of persistence reflects changes in spatial locations between the two time periods (Figure 4, *left and middle*). During 1926–1969, persistent droughts (Figure 4; *left*) extend across the Southern Great Plains, Midwest and West.  In the next half, droughts became persistent over the West, Southwest, parts of the Midwest and Appalachia. The difference map (Figure 4; *right*) shows an intensification of persistent droughts over the Southwest, South, Central, Southeast, and part of the Midwest in the next 44 years. However, changes in the persistence probability are limited to only ~ 3% of the stations and are not field significant.

The spatial patterns of seasonal drought persistence (Figure S6) show strong seasonal variations overall in the CONUS. Summer-Fall droughts are persistent over the Northern Plains and parts of the Midwest, Southeast and Central US. On the contrary, Winter-Spring persistent droughts are limited to the Northwest and portions of the Southeast. Locally significant ($\alpha_L = 0.10$) growth and decrease in Summer-Fall persistent droughts are limited to less than 5% of the stations. Roughly an equal number (around 2%) of stations shows growth and decrease in Winter-Spring persistent droughts. No emergence of persistent droughts is observed either in the Summer-Fall or in the Winter-Spring seasons. Trends in seasonal drought persistence are not field significant ($\alpha_f = 0.10$). In any case, seasonal droughts cannot adequately represent multi-year persistence. Our analyses suggest that barring a few regional exceptions, the overall drought persistence has been stationary over time, irrespective of the seasons and time frames (such as multi-year drought episodes) considered.

*3.3 Spatial Variability in Regional Drought Severity*

Figure 5 (*left*) shows the spatial distribution of the average (weighted by duration) drought severity (*left*) during 1970–2013. Average drought duration during 1970–2013 varies from 2 to 5 months with a coefficient of variation of 0.15. Based on the Shaprio-Wilk test (at α = 0.05) (Shapiro and Wilk, 1965), the normality assumptions of drought properties (severity and duration) are rejected for all stations, following which the correlation between drought properties are assessed using Kendall's τ rank correlation (a non-parametric approach). The spatial plot of the correlation patterns between drought severity and duration is shown in Figure 5 (*right*). Correlations are found to be field significant (α = 0.05) in all regions and about 11% of the stations show strong dependence (Kendall's τ > 0.9). The strength of dependence between drought severity and duration can arise in four different cases as shown in Figure S7. The dependence pattern can be relatively weak owing to the presence of high (low) values of severity with short (long) durations (as shown by event 1 and event 2 in Figure S7 *top panel*). Conversely, dependence is stronger when a particular event is characterized by high (low) severity value with long (short) duration (as shown in event 3 and event 4 in Figure S7; *bottom panel*). For example, the spatial dependence of severity-duration increases with higher average severity and vice versa. Therefore, dependence between drought properties cannot be ignored in the development of SDF curves and in the corresponding translation to risk management.

Spatial distribution plots of severity for $T = 10$–year and $T = 100$–year return periods are generated for individual stations using the SDF relation, conditional on drought duration up to 12 months (Figure S8; *left and right panel*) respectively. After calculating drought severity conditional to the duration for each station, the high, low and mean values corresponding to each region are derived using a percentile-based approach. The SDF-derived severity values from individual stations in a region are ranked in descending order and the 50th, 10th and 90th percentile values are calculated for each month, providing a regional measure of the central tendency as well as the lower and upper (inter-quantile) bounds. The SDF curves for the entire CONUS region and each of the nine NCDC climate regions are shown in Figure 6 (*left and right panel*) for the recent period (1970–2013). The spatial pattern (in Figure S8) shows intensification of severe droughts over Northern California that extends all the way to the Northwest, as well as parts of the Southwest, West-north Central, South, Southeast and East-north Central regions under moderately extreme conditions (10-year return period). Under the most extreme conditions (100-year return period), most of the regions show increasing severe droughts, including most of the West, Northwest, South, Southeast and portions of the Southwest and West-north Central, as well as the Northeast and Midwestern regions respectively.

Based on severity values derived from SDF relations, changes in the medians and slopes of drought severity of extreme droughts (10- and 100-year return periods) are found over time (Figure 7). The spatial plots of the median and slope of drought severity show a lack of clear spatially coherent patterns. The spatial distribution plot of severity, together with the regional SDF curves may inform agro-meteorological planning; as an example, the yields of certain crops are expected to reduce if an event exceeds specific severity or duration thresholds (Basso and Ritchie, 2014). The SDF curves reveal variations in regional drought severities, which in turn may be useful for drought monitoring and designing water supply or storage systems to prepare against severe droughts.

*3.4 Stationarity in Variance of Extremes with Increasing Return Period*

Changes in the variance of average drought severity show mixed upward/downward field significant trends over most of the regions except in the Southwest. Over the Central (16% of stations; 15 out of 92 stations in Central region) and Northeast (21% of stations; 14 out of 67 stations in Northeast region) regions, a strong downward trend is observed (Table S7). However, the proportions of stations with insignificant trends outnumber stations with up/downward trends. We observe no field significant nonstationarity in variance for the spatial coverage of drought extremes (Figure 8). A similar pattern is observed in the spatial variability of extremes without considering any outlying years (Figure S9).

**4. Discussion of Results and Plausible Mechanistic Interpretations**

Our results reveal that despite uncertainty in the trends of the spatial coverage of mean meteorological drought severity, the corresponding extremes have been exhibiting increasing trends across multiple regions in the conterminous United States. In addition, over the last several years, the coverage of extreme droughts exceeds the coverage in the dust bowl era. We hypothesize that the elevated extreme drought risks over North America may be linked to the role of atmospheric variability on regional drought, which has also been suggested by previous studies. For example, a recent study (Kam et al. 2014) shows that the coupling effect of the Pacific Decadal Oscillation (PDO) and the El Niño-Southern Oscillation (ENSO) are responsible for the increase in annual meteorological drought risk over the Southern US. The authors (Kam et al. 2014) further suggest that in recent decades the strength of the coupling effect has weakened and shifted to the Southwestern US. Likewise, the severe and spatially variable North-West droughts are attributed to the presence of Pacific Blocking off the Northwest coast (Knapp et al., 2004). Over the Northeast it has been shown that increases in evapotranspiration owing to an extension of the growing seasons could intensify the frequency of drought (Huntington et al., 2009). In addition, using observations and climate model experiments, (Groisman et al., 2005) show an increase in the number of dry spells in recent decades, which is primarily accompanied by an increase in precipitation totals with a decrease in the number of rainy days over the Northeast. Based on analysis of 1951–2010 data, a recent study suggests that annual evapotranspiration rates have increased considerably over much of the US, with significant increases in the Southwest and Southeast (Jung et al., 2013).

We find that the variance of extreme drought severity remains stationary irrespective of the data and time periods considered although there are changes in the average of these extremes. A similar insight regarding changes in the mean but no change in variance has been noted previously by other researchers for summer temperature extremes (Rhines and Huybers, 2013). Our analysis shows that drought persistence remains field insignificant. However, we find evidence of locally significant drought persistence increase over Southeast, Central and Southwestern United States. Increased variability in summer precipitation over the Southeast has been found to be closely linked to the intensification and westward shift of North Atlantic subtropical high [*NASH*; (Li et al., 2011)], whereas dry winter is weakly associated with La Niña conditions in the tropical Pacific Ocean (Seager et al., 2009). While the influence of large-scale circulation patterns is largely responsible for the occurrence of persistent droughts over North America (McCabe et al., 2004), the increase in spatial coverage of extremes may pose a critical challenge for drought preparedness and monitoring. Finally, global warming has been suggested as a plausible explanation for the increase in drought severity trends (Cayan et al., 2010; Karl et al., 2009; Seager et al., 2012, 2009; Seager and Vecchi, 2010; Strzepek et al., 2010).

## 5. Summary of Key Results and Conclusions

This paper explores an urgent and relevant question in drought climatology over the conterminous US: *Is spatial coverage of severe droughts becoming more in recent times and are the trends in extreme droughts different from overall mean trends in different regions*? We present concurrent insights for mean and extreme trends in US droughts through comprehensive analyses based on two observational records. These patterns, while not apparent from standard hydrological data analyses, may be critical for extreme drought preparedness and monitoring.

The key insights are summarized as follows:

- Spatial coverage of the severity of extreme meteorological droughts over drought-prone regions of the CONUS exhibit an increasing trend, and this trend in turn is robust to the selection of datasets and to the inclusion or exclusion of the outlying drought years.

- Mean meteorological drought severity is more sensitive to the choice of datasets than to the presence or absence of outlying drought years.

- The trends identified in extremes (10-year and 100-year) are sensitive to outlying years but consistent across datasets.

- The persistence property of droughts remains relatively stationary across regions.

- Trends in the temporal variance of average drought severity show considerable regional variability in field significance while the temporal variance of extreme droughts remains stationary regardless of the data used and the time frames considered.

- The paper presents proof of principle results which suggest that copula-based SDF curves can be designed for droughts to offer quantitative guidance to stakeholders and planners.

Several caveats should be considered. While meteorological droughts are occasionally precursors for potentially more damaging agricultural or hydrological droughts, any generalization needs to be made with caution. For example, precipitation-based indices may not be able to capture snow-related events especially in the western US (Cayan et al., 2010; Pederson et al., 2011). Likewise, soil moisture, which is key to the more damaging droughts, exhibits different persistence properties (Cook et al., 2007; Seager et al., 2005) compared to precipitation. In the future, there is a need to explore trends in mean and extreme droughts by considering the multivariate influence of precipitation and temperature (AghaKouchak et al., 2014; Diffenbaugh et al., 2015). Finally, while we have validated each of our findings carefully by rigorous statistical tests, including field significance tests, the results are nevertheless contingent on the sample size and observed data quality.

**Acknowledgments:**

This work was primarily funded by two National Science Foundation grants, specifically, the Big Data grant # 1447587 entitled "High-Dimensional Statistical Machine Learning for Spatio-Temporal Climate Data" and the Expeditions in Computing Grant # 1029711 entitled "Understanding Climate Change: A Data-driven Approach". Northeastern University provided partial funding. Station-based hourly (NCDC DSI-3240) and monthly (United States Historical Climatology Network; USHCN Version 2.5) precipitation data are obtained from the website of the Carbon Dioxide Information Analysis Center (CDIAC), Oak Ridge National Laboratory and the National Climatic Data Center (NCDC) respectively. The first author of the manuscript would like to thank Dr. Benjamin Renard of the IRSTEA, France and Prof. Daniel Wilks of the Cornell University for their fruitful suggestions on implementation of FDR-based field significance test. The authors would like to thank Dr. Vimal Mishra of IIT-Gandhinagar in India, who was a visiting faculty at the SDS Lab at Northeastern, for his help with access to data from only those stations which do not have missing values. The authors thank Dr. Evan Kodra, Dr. Daiwei Wang, Mr. Udit Bhatia and Mr. Venkata Shashank Konduri, former or current graduate students at the Sustainability and Data Sciences Lab of Northeastern University, for helpful comments and suggestions, as well as Frank Capogna, a graduate student in the English Department and a member of Northeastern University's writing center, for helpful editorial comments.

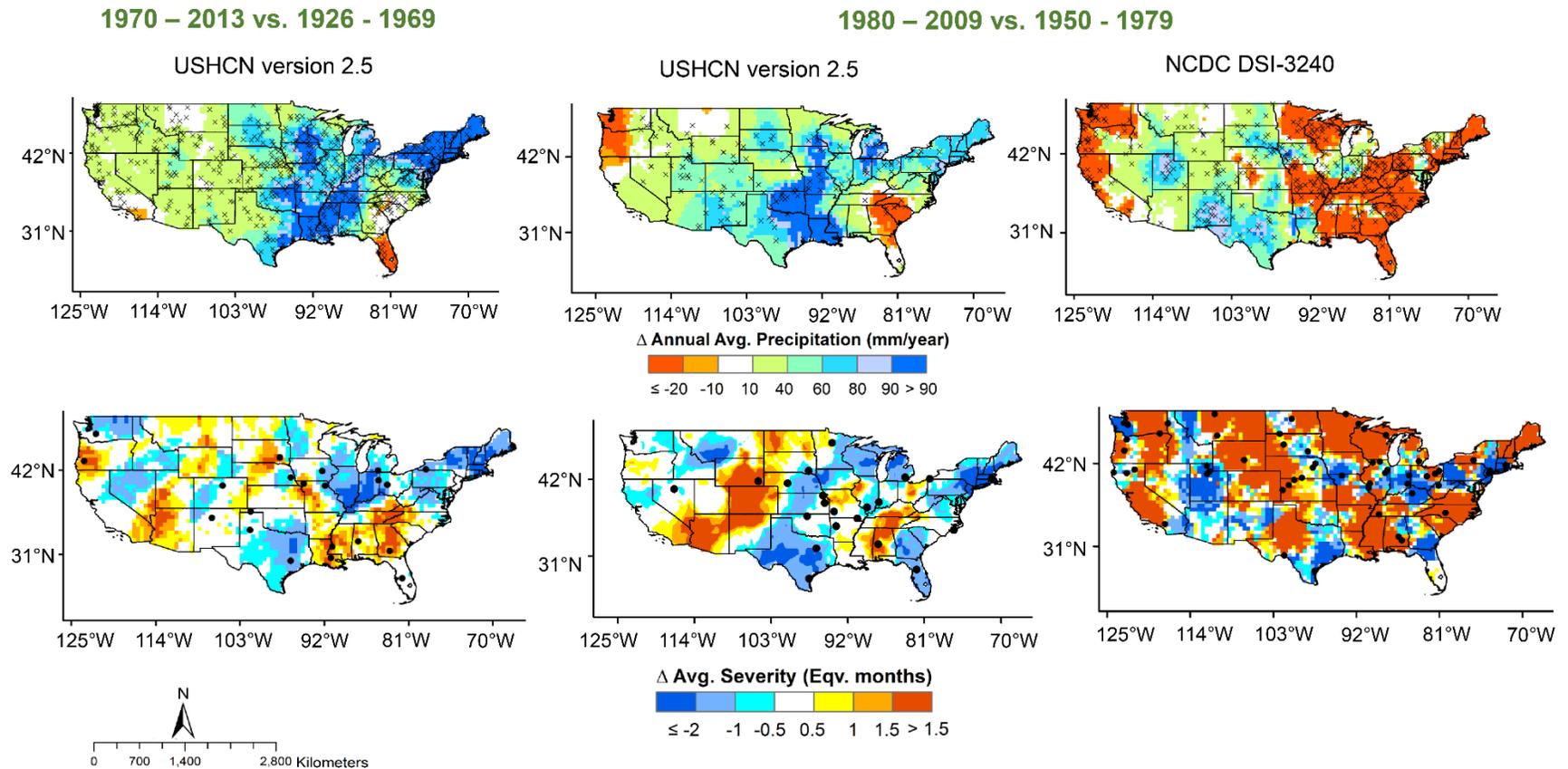

**Fig.1.** Changes in annual average precipitation (*top*) and severity (*bottom*) for 1970 – 2013 versus 1926 – 1969 (*left*) in D1-Out (dataset 1 with outlying years), and 1980-2009 versus 1950 – 1979 (*middle*) in D1-No Out (dataset 1 without outlying years) and (*right*) D2-No Out (dataset 2 without outlying years). Stipples (in difference map of annual average precipitation) and filled black circles (in difference map of average severity) indicate locally significant trends at 5% significance level.

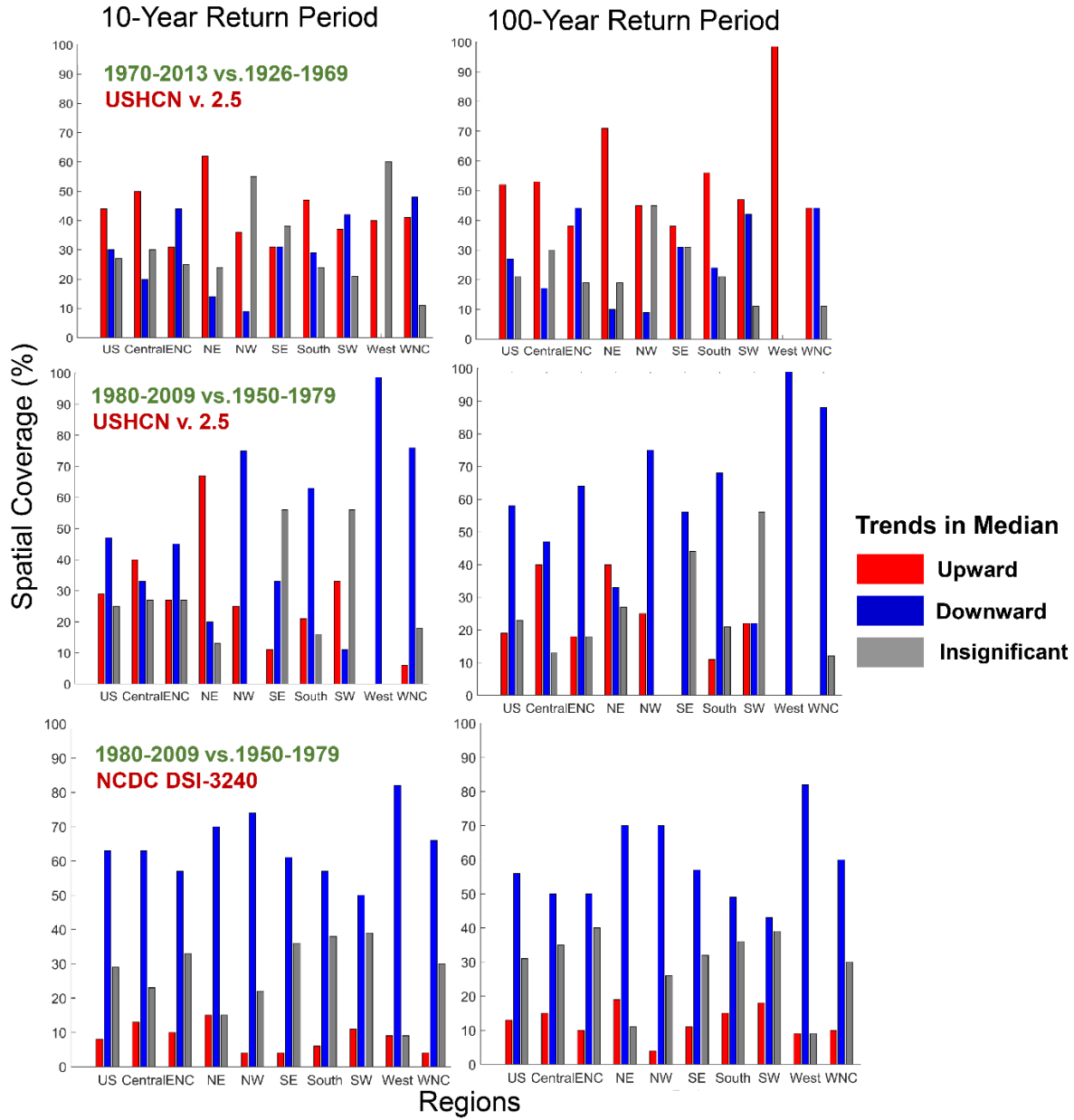

**Fig. 2.** Spatial coverage of changes in median trends in severity of extreme drought events corresponding to 10- (*left*) and 100- (*right*) year return periods for 1970-2013 versus 1926-1969 (*top*) in D1-Out dataset, and 1980 – 2009 versus 1950 – 1979 (*middle*) in D1-No Out and (*bottom*) D2-No Out.

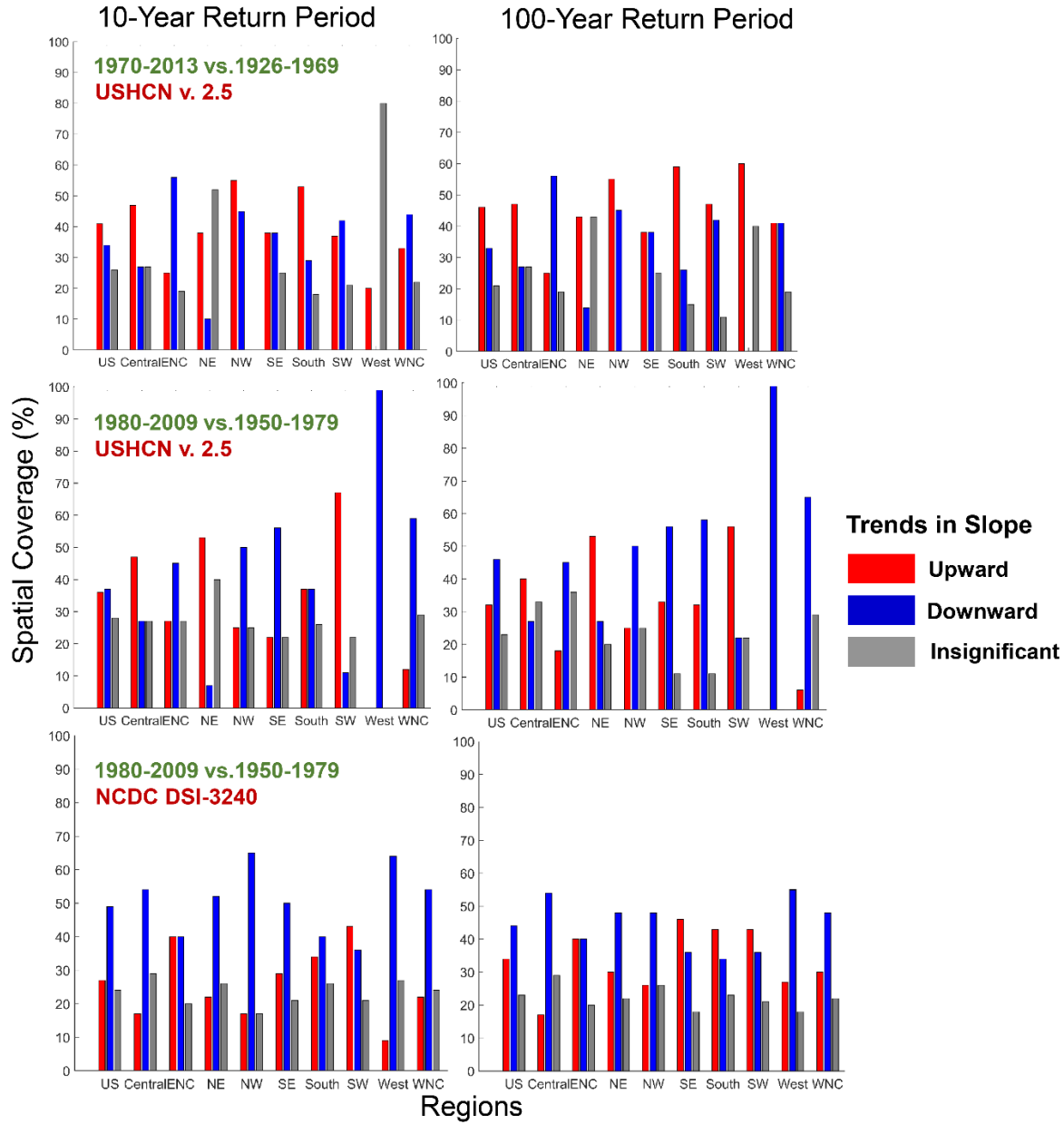

**Fig. 3.** Spatial coverage of changes in slope in severity of extreme drought events corresponding to 10- (*left*) and 100- (*right*) year return periods for 1970-2013 versus 1926-1969 (*top*) in D1-Out dataset, and 1980 – 2009 versus 1950 – 1979 (*middle*) in D1-No Out and (*bottom*) D2-No Out.

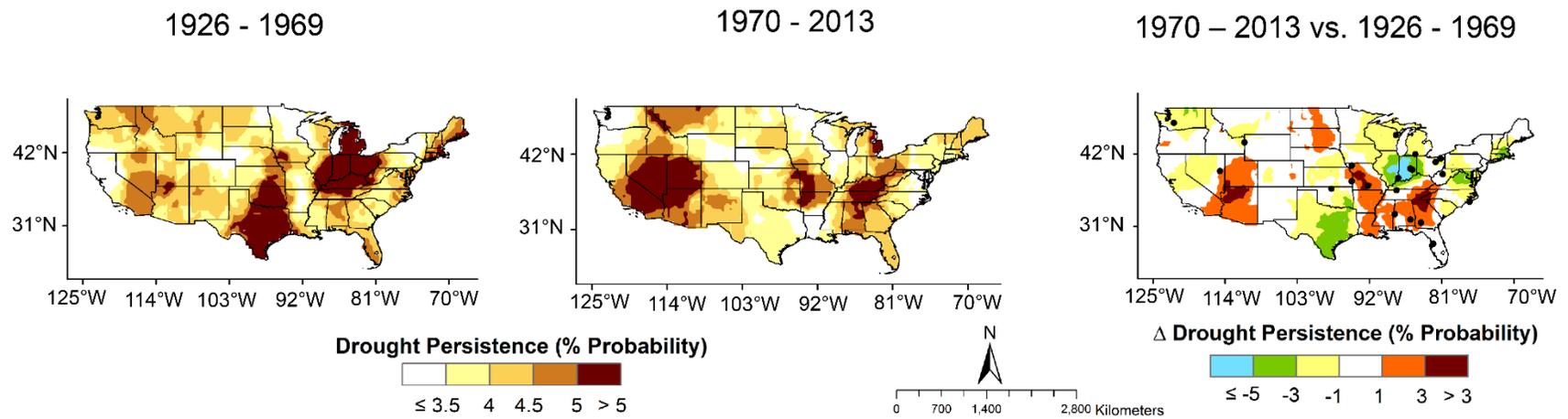

**Fig. 4.** Persistence probability of droughts in two consecutive time-windows 1926-1969 (*left*), 1970-2013 (*middle*), and the corresponding difference map (*right*) comparing 1970 – 2013 versus 1926-1969. Locally significant ($\alpha_L = 0.10$) differences in persistence probability are marked with filled black circles.

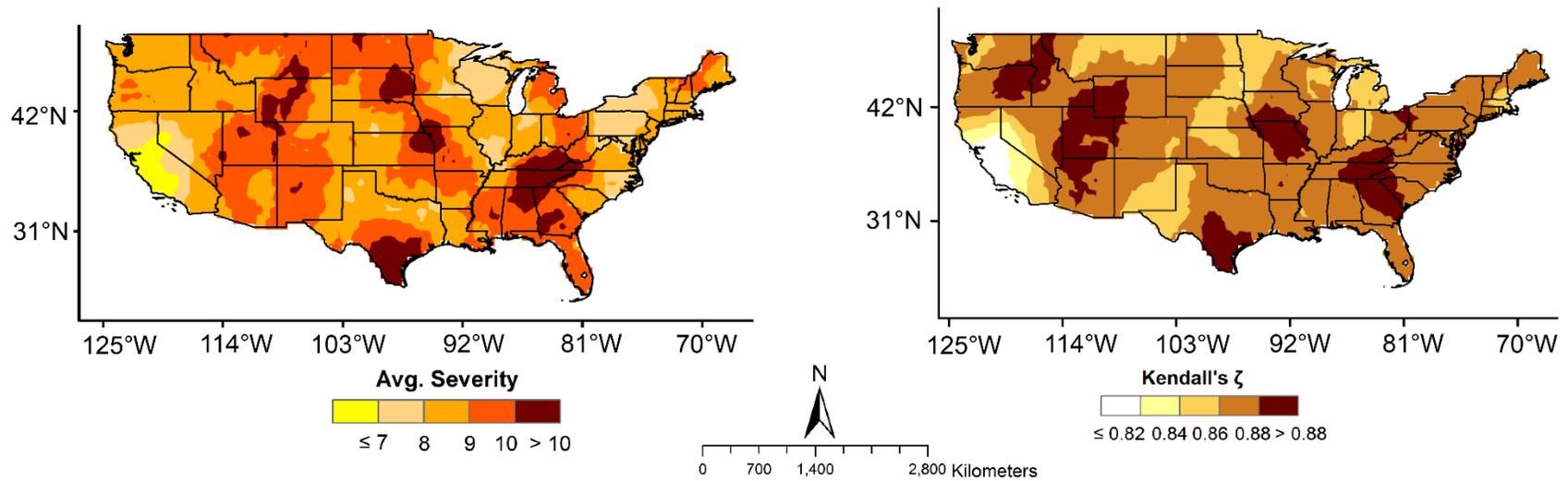

**Fig. 5.** Spatial distributions of (*left*) weighted average severity (in equivalent months), where weights are calculated based on drought duration, and (*right*) spatial dependence patterns (using Kendall's τ correlation) during 1970-2013.

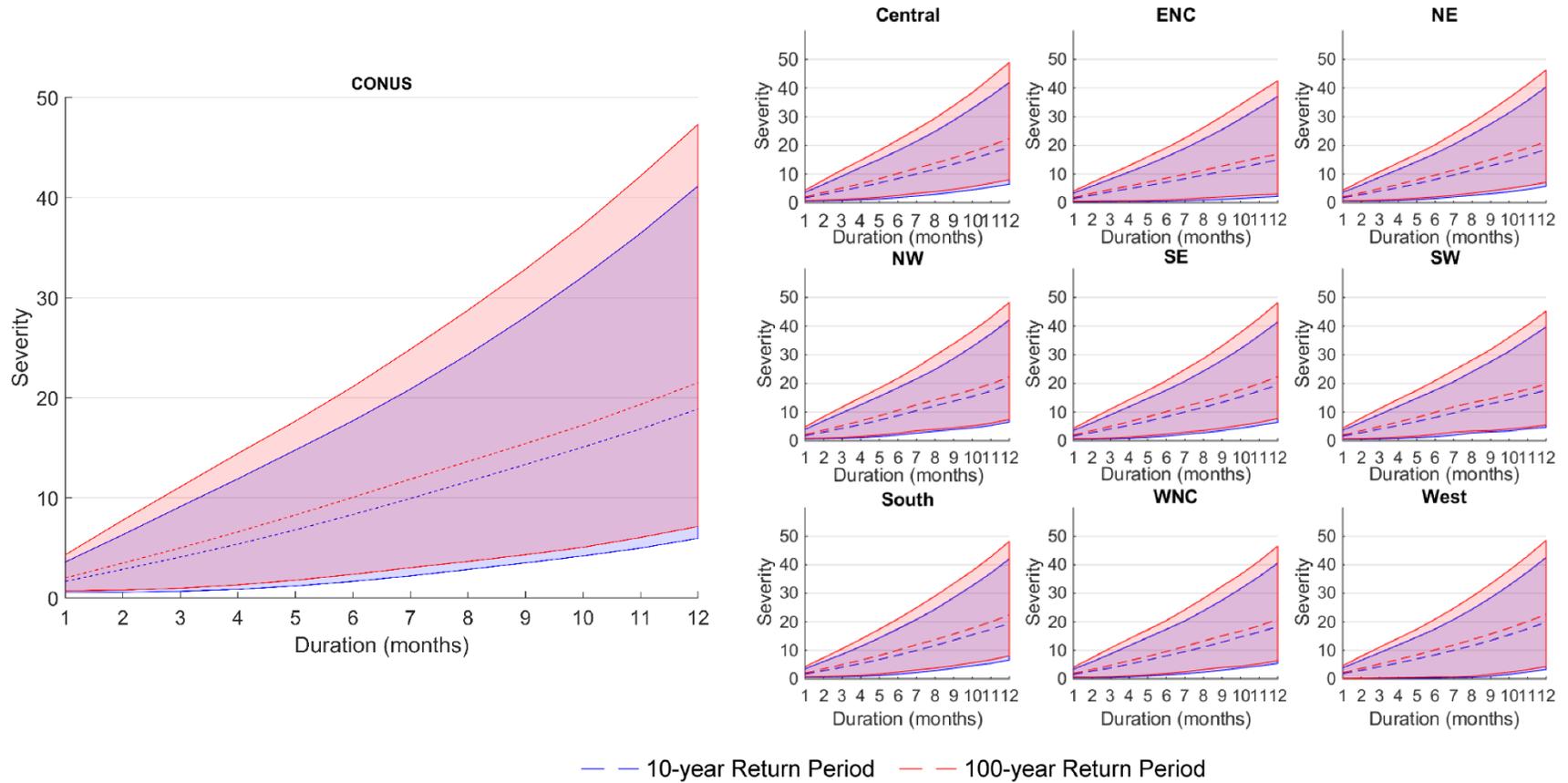

**Fig. 6.** Drought Severity-Duration-Frequency (SDF) curves at 10- and 100-year return periods for (*left*) the CONUS in aggregate and (*right*) the nine NCDC climate regions. The median (50th percentile) drought severity values from individual stations are used for the construction of regional SDF curves, whereas the shaded regions indicate upper (90th percentile) and lower (10th percentile) bounds of severity values corresponding to each region.

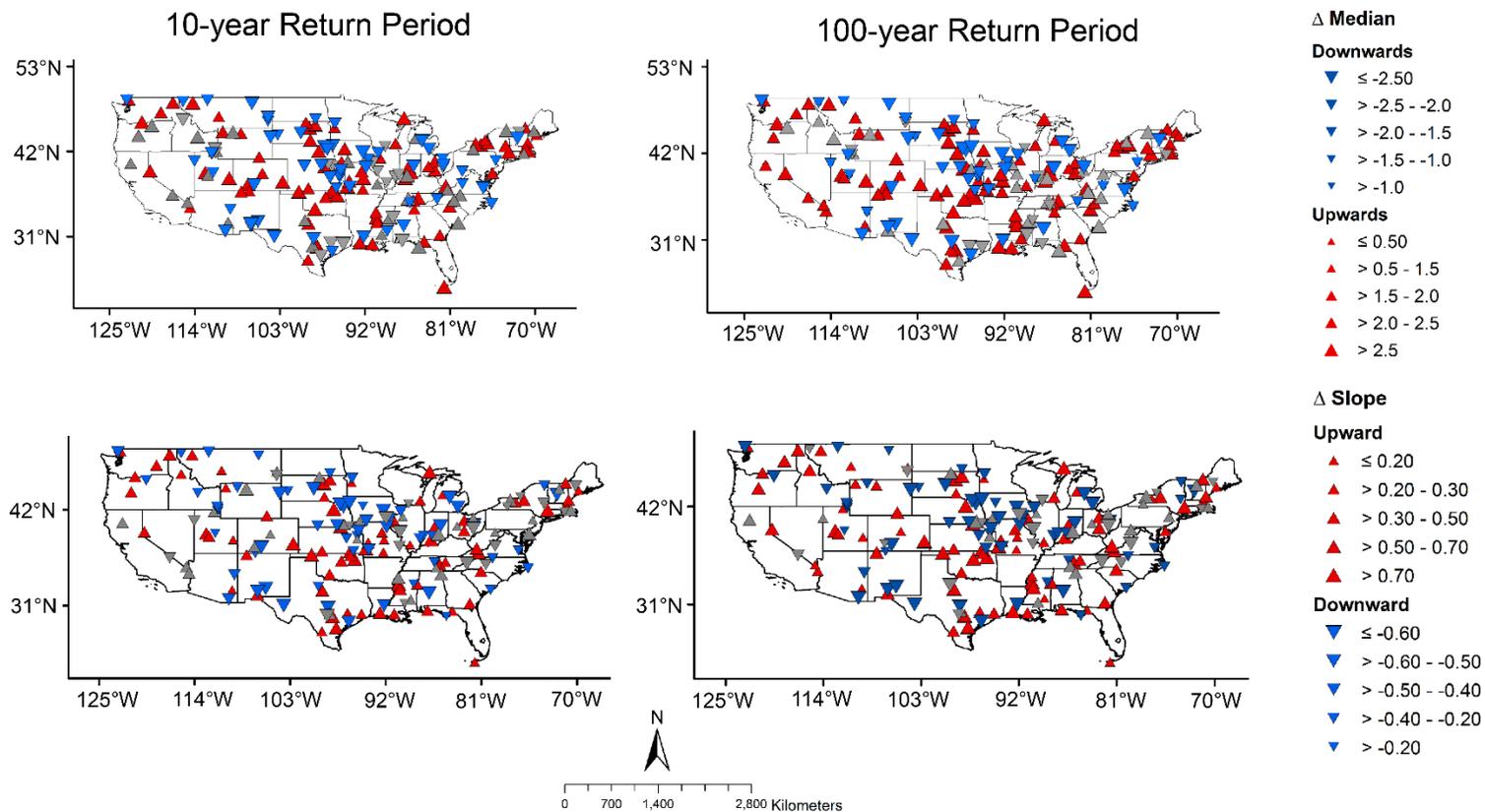

**Fig. 7.** Changes in median (*top*), and slope (*bottom*) of, severity at 10- (*left*) and 100-year (*right*) return periods. Geographical locations of the stations are shown as triangles while the colors describe the sign and the significance of the estimated trend. Upward, downward and insignificant trends are marked with red, blue and gray triangles respectively. The size (and shading) of the triangle is proportional to the magnitude of the trend. The legend indicating changes in median and the slope applies to both panels.

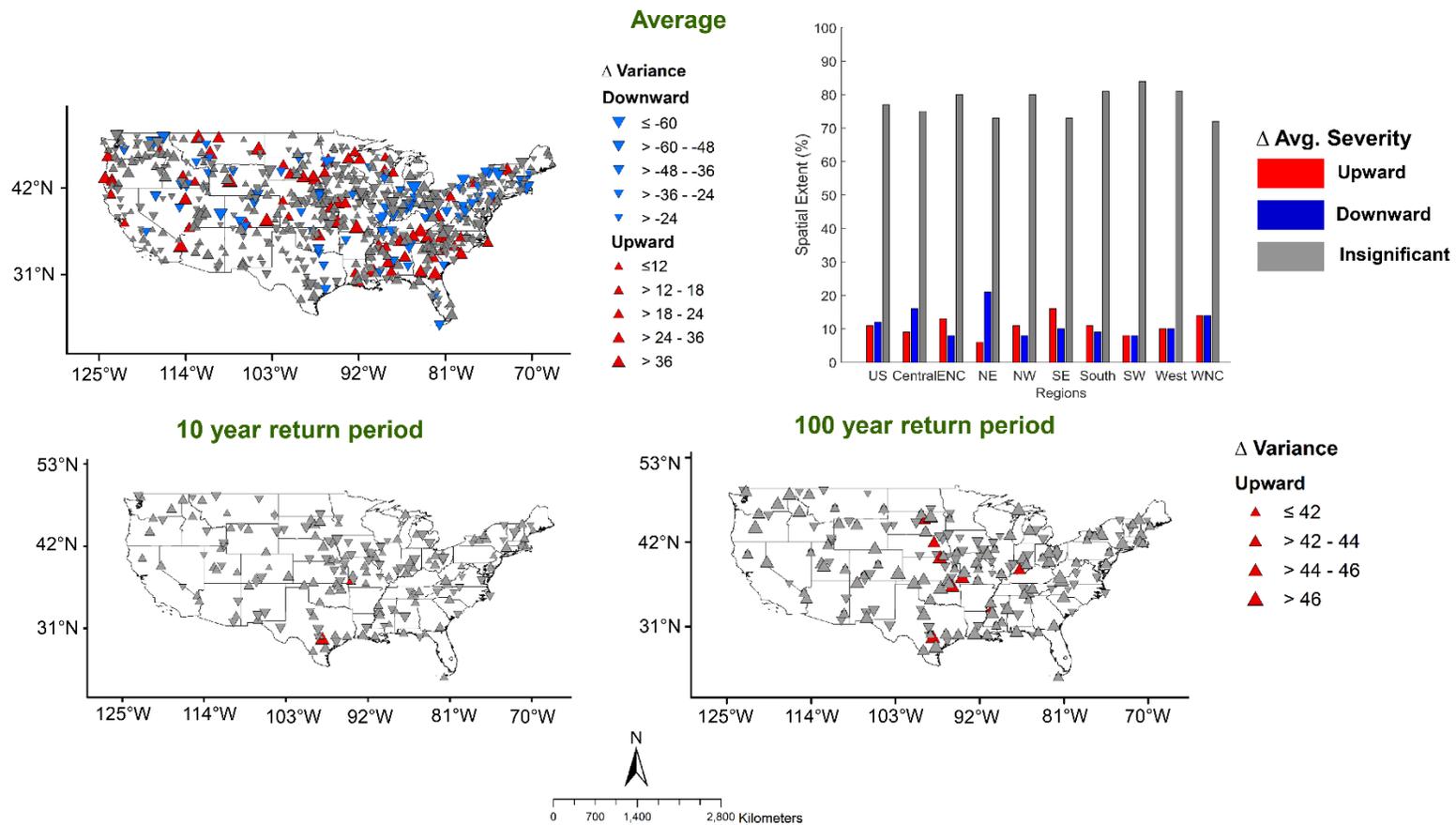

**Fig. 8.** Changes in the (temporal) variance of average (*top*) and extreme (*bottom panel*) drought severity. The top panel (*right*) shows spatial coverage, specifically those with locally significant trends, in average drought severity. The drought severity associated with 10- and 100-year return periods are shown as the extreme drought categories. Triangles indicate geographical locations of the stations and the colors describe the sign and the significance of the estimated trends. The upward, downward and insignificant trends are marked with red, blue and gray colored triangles. The size (and shading) of the triangle is proportional to the magnitude of the trend.

Supplemental Material for

# Space-time Trends in U.S. Meteorological Droughts

Poulomi Ganguli[1] and Auroop R. Ganguly[1]*

[1]Sustainability and Data Sciences Laboratory, Civil and Environmental Engineering, Northeastern University, 360 Huntington Avenue, Boston, MA, 02115, USA, *Correspondence to: *poulomizca@gmail.com*

**Contents of this file**

1. Tables S1 to S7

2. Figures S1 to S9

**Introduction**

This supporting information provides ancillary tables and figures.

**Table S1.** Current understanding of spatiotemporal pattern of dry spells in the CONUS

| Trend | Insights | Region | Time Period | Timescale | Index | Reference |
|---|---|---|---|---|---|---|
| Drought | Severity and frequency of drought tends to decrease in post 1935-36, however (insignificant) trend toward wetter conditions in the South and drier conditions in the West north Central region was evident. | CONUS[1] | 1895 – 1989 | Monthly | PHDI[2] | Karl and Heim (1990) |
| | Shift towards wetter condition in 38 out of 48 states during 1955 – 1989 as compared to 1895 – 1954. | CONUS | 1895 – 1989 | Monthly | PDSI[3] | Idso and Balling Jr (1992) |
| | Spatial patterns of drought frequency and duration are largely influenced by the index used to define drought. | CONUS | 1895 – 1988 | Monthly | ZINX[4], PDSI, PHDI | Soulé (1992) |
| | Comparisons between the early and middle 30-year periods show transition from wetter to drying pattern in the regions stretching from Northcentral Rocky Mountains to the northern Great Plains. | CONUS | 1900 – 1989 | Monthly | PHDI | Soulé (1993) |
| | Positive mean trends for 95, 50 and 30 year time windows; negative for 15 year time window indicates perception of trends is partially function of temporal scaling. | CONUS | 1895 – 1989 | Monthly | PHDI | Soulé and Yin (1995) |
| | Except a few decreasing trend in part of Southwest, overall an increasing trend for both modeled soil moisture and runoff, indicating shift towards the wetter condition. | CONUS | 1915 – 2003 | Monthly | VIC[5] simulated soil moisture and runoff at 0.5° spatial resolution | Andreadis and Lettenmaier (2006) |
| | Except West and Southwest, the observed increase in precipitation since 1980s', masked the tendency of increasing drought during 20th century. | CONUS | 1950 – 2006 | Monthly | PDSI | Easterling et al. (2007) |
| | Increase in mean duration of dry episodes in Eastern and Southwestern regions | CONUS | 1961 – 2006 | Daily | Station observed precipitation data | Groisman and Knight (2008) |
| | More intense summer drought and anomalous wetness during 1978-2007 as compared to 1948-1977. | SEUS[6] | 1948-2007 | Daily, Monthly | Precipitation anomaly, SPI – 3, SPI - 9 | Wang et al. (2010) |
| | Increased drought intensity for many areas in eastern SUS (portion of North and South Carolina), although no significant changes in drought area and duration in most region. | SUS[7] | 1895 – 2007 | Monthly | SPI[8] – 1, SPI – 3, SPI – 6, SPI – 12 | Chen et al. (2012) |
| | Overall tendency towards wetness | CONUS | 1900-2002 | Annual | PDSI[*] | Dai et al. (2004) |
| | Mixture of dry and wet pattern with tendency towards wetness in most of the regions; signature of drying over part of Northwest and West | CONUS | 1950-2008 | Annual average | PDSI – Penman-Monteith formulation | Sheffield et al. (2012) |
| Drought persistence | Droughts persist much longer in the interior part of CONUS as compared to the regions close to the coasts. | CONUS | 1895 – 1988, 1948 – 2004 | Monthly | ZINX, PDSI, PHDI, SPI | Soulé (1992); Mo and Schemm, (2008) |
| | Drought in Northeast is less persistent but may persist more over Northwest and Southwest of US due to influence of low frequency decadal SST[9] variability. | CONUS | 1948 – 2004 | Monthly | PDSI, SPI-6, SPI-12, SPI-60 | Mo and Schemm, (2008) |
| | Seasonal drought persistence over North-Central Alabama is more as compared to the other regions. | SEUS[6*] | 1895 – 2011 | Monthly | SPI-3 | Ford and Labosier, (2013) |
| Dry days | Increase in precipitation totals with decrease in number of rainy days, indicating increase in number of dry days. | NEUS[10] | 1961 - 1990 | Daily | Station observed precipitation data | Groisman et al. (2005) |
| | Decrease in mean consecutive dry days since 1960s. | CONUS | 1951 – 2003 | Daily | Station observed temperature and precipitation data | Alexander et al. (2006) |
| | Decrease in precipitation frequency and increase in dry spell length during wet season in Southeast despite increase in precipitation extremes accompanied by decrease in dry spell length in most of the regions. | CONUS | 1930-2009 | Daily | Station observed precipitation data | Pal et al. (2013) |

*Note:*[1]CONUS: Conterminous United States, [2]PHDI: Palmer Hydrological Drought Index, [3]PDSI: Palmer Drought Severity Index, [4]ZINX: Palmer Moisture Anomaly Index, [5]VIC: Variable Infiltration Capacity, [6]SEUS: Southeastern US (25° - 36.5°N, 76° - 91° W), SEUS[6*]: Southeastern US including Texas and Oklahoma; [7]SUS: Southern US (including Mid-Atlantic coast West to Texas), [8]SPI: Standardized Precipitation Index, [9]SST: Sea Surface Temperature, [10]NEUS: Northeastern US

[*] Evapotranspiration component in PDSI is calculated using a simple scheme without taking into account the effects of changes in surface solar radiation, relative humidity and wind speed.

Table S2. Field significance tests for trends in average annual precipitation and average severity

| Region | D1-Out 1970-2013 vs. 1926-69 | | | | D1-No Out 1980-2009 vs. 1950-1979 | | | | D2-No Out 1980-2009 vs. 1950-1979 | | | |
|---|---|---|---|---|---|---|---|---|---|---|---|---|
| | Upward | | Downward | | Upward | | Downward | | Upward | | Downward | |
| | $p_{fdr}$ | $N$ | $p_{fdr}$ | $N$ | $p_{fdr}$ | $N$ | $p_{fdr}$ | $N$ | $p_{fdr}$ | $N$ | $p_{fdr}$ | $N$ |
| Central | 0.014 | 26 | 0.005 | 0 | $1.27e^{-4}$ | 0 | 0 | 0 | 0.011 | 18 | 0.021 | 57 |
| ENC | 0.020 | 24 | 0 | 0 | 0.001 | 1 | 0 | 0 | 0.014 | 13 | 0.020 | 40 |
| Northeast | 0.029 | 42 | - | - | 0.004 | 5 | 0 | 0 | 0.013 | 8 | 0.028 | 36 |
| Northwest | 0.001 | 2 | 0 | 0 | $4.6e^{-4}$ | 1 | $3.12e^{-4}$ | 0 | 0.012 | 7 | 0.018 | 27 |
| Southeast | 0.002 | 3 | 0 | 0 | $3.7e^{-8}$ | 0 | 0 | 0 | 0.001 | 1 | 0.018 | 29 |
| South | 0.010 | 16 | 0 | 0 | 0.003 | 7 | 0 | 0 | 0.008 | 13 | 0.009 | 11 |
| Southwest | 0.014 | 12 | 0.010 | 2 | 0.024 | 22 | 0.0014 | 0 | 0.026 | 42 | 0.006 | 4 |
| West | 0.006 | 2 | 0 | 0 | $1.61e^{-4}$ | 0 | 0 | 0 | 0.004 | 4 | 0.026 | 28 |
| WNC | 0.014 | 17 | 0.003 | 1 | 0.004 | 5 | 0.004 | 1 | 0.011 | 28 | 0.006 | 7 |

*Note:* [1]$p_{fdr}$ denotes *p*-value threshold that controls the FDR at α = 0.05 significance level, *N* denotes the number of sites with *p-values* less than $p_{fdr}$, $N \geq 1$ indicates the regional trend is field significant; 0 values in $p_{fdr}$ column indicate threshold does not exist.

[2]For average severity, the D2-No Out data showed field significant downward trend in NE region. Trends in average severity in other regions are not field significant in this dataset. On the other hand, none of the regions show field significant trend for the changes in average severity for USHCN dataset (*i.e.*, D1-Out and D1-No Out).

| Dataset | Station | NCDC Region/State | Longitude | Latitude | Analysis period | No. of Events | Severity | | | Duration | | |
|---------|---------|-------------------|-----------|----------|-----------------|---------------|----------|--|--|----------|--|--|
| | | | | | | | Distribution | $d_{KS}$ | $d_{KS}^{\alpha=0.05}$ | Distribution | $d_{KS}$ | $d_{KS}^{\alpha=0.05}$ |
| D1-Out | 1 | East-North Central (Iowa) | -92.28 | 41.88 | 1926 – 1969 | 28 | Lognormal | 0.08 | 0.71 | Exponential | 0.14 | 0.19 |
| | 2 | West (California) | -116.87 | 36.46 | 1926 – 1969 | 30 | Lognormal | 0.14 | 0.69 | Exponential | 0.14 | 0.18 |
| | 3 | Central (Tennessee) | -82.98 | 36.42 | 1970 – 2013 | 35 | Gamma | 0.11 | 0.15 | Exponential | 0.08 | 0.18 |
| | 4 | Northeast (Massachusetts) | -71.11 | 42.21 | 1970 – 2013 | 27 | Weibull | 0.11 | 0.15 | Exponential | 0.10 | 0.19 |
| D1-No out | 1 | East-North Central (Iowa) | -92.28 | 41.88 | 1950 – 1979 | 17 | Lognormal | 0.10 | 0.73 | Exponential | 0.10 | 0.23 |
| | 2 | West (California) | -116.87 | 36.46 | 1950 – 1979 | 26 | Lognormal | 0.15 | 0.68 | Lognormal | 0.31 | 0.68 |
| | 3 | Central (Tennessee) | -82.98 | 36.42 | 1980 – 2009 | 20 | Lognormal | 0.07 | 0.73 | Exponential | 0.12 | 0.21 |
| | 4 | Northeast (Massachusetts) | -71.11 | 42.21 | 1980 – 2009 | 18 | Lognormal | 0.09 | 0.74 | Exponential | 0.12 | 0.23 |
| D2-No out | 1 | East-North Central (Iowa) | -95.44 | 41.51 | 1950 – 1979 | 14 | Weibull | 0.16 | 0.22 | Exponential | 0.15 | 0.25 |
| | 2 | West (California) | -117.93 | 33.89 | 1950 – 1979 | 29 | Lognormal | 0.08 | 0.74 | Exponential | 0.16 | 0.18 |
| | 3 | Central (Illinois) | -91.01 | 41.42 | 1980 – 2009 | 24 | Lognormal | 0.11 | 0.70 | Exponential | 0.11 | 0.19 |
| | 4 | Northeast (New Hampshire) | -70.95 | 43.15 | 1980 – 2009 | 11 | Lognormal | 0.15 | 0.77 | Exponential | 0.16 | 0.28 |

*Note*: $d_{KS}$ denote distance between theoretical and empirical distribution by Kolmogorov-Smirnov (K-S) test statistics. $d_{KS}^{\alpha=0.05}$ is the critical value of K-S test statistics obtained by $n = 1000$ bootstrapped iterations.

**Table S4.** Goodness-of-fit tests of copula families for a few selected meteorological stations

| Dataset | Station | NCDC Region/State | Longitude | Latitude | Analysis period | No. of Events | Kendall's $\tau_{s,d}$ | Copula Family | Parameter | $p_{value}$ |
|---------|---------|-------------------|-----------|----------|-----------------|---------------|------------------------|---------------|-----------|-------------|
| D1-Out | 1 | West (California) | -116.87 | 36.46 | 1926 – 1969 | 30 | 0.892 | Frank | $\theta = 12.71$ | 0 |
| | | | | | | | | Gumbel | $\theta = 4.36$ | 0 |
| | | | | | | | | Plackett | $\theta = 79.41$ | 0.004 |
| | | | | | | | | Student's $t$ | $\vartheta = 6,\ \rho_{s,d} = 0.988$ | 0.064 |
| | 2 | Central (Tennessee) | -82.98 | 36.42 | 1970 – 2013 | 35 | 0.883 | Frank | $\theta = 14.59$ | 0.008 |
| | | | | | | | | Gumbel | $\theta = 3.96$ | 0 |
| | | | | | | | | Plackett | $\theta = 55.113$ | 0 |
| | | | | | | | | Student's $t$ | $\vartheta = 6,\ \rho_{s,d} = 0.987$ | 0.316 |
| D1-No out | 1 | West (California) | -116.87 | 36.46 | 1950 – 1979 | 26 | 0.834 | Frank | $\theta = 7.08$ | 0 |
| | | | | | | | | Gumbel | $\theta = 2.88$ | 0 |
| | | | | | | | | Plackett | $\theta = 25.97$ | 0 |
| | | | | | | | | Student's $t$ | $\vartheta = 6,\ \rho_{s,d} = 0.994$ | 0.036 |
| | 2 | Central (Tennessee) | -82.98 | 36.42 | 1980 – 2009 | 20 | 0.894 | Frank | $\theta = 15.106$ | 0.048 |
| | | | | | | | | Gumbel | $\theta = 4.30$ | 0.028 |
| | | | | | | | | Plackett | $\theta = 59.37$ | 0.008 |
| | | | | | | | | Student's $t$ | $\vartheta = 6,\ \rho_{s,d} = 0.991$ | 0.44 |
| D2-No out | 1 | West (California) | -117.93 | 33.89 | 1950 – 1979 | 29 | 0.869 | Frank | $\theta = 10.70$ | 0 |
| | | | | | | | | Gumbel | $\theta = 3.35$ | 0 |
| | | | | | | | | Plackett | $\theta = 30.25$ | 0 |
| | | | | | | | | Student's $t$ | $\vartheta = 8,\ \rho_{s,d} = 0.979$ | 0.136 |
| | 2 | Central (Illinois) | -91.01 | 41.42 | 1980 – 2009 | 24 | 0.872 | Frank | $\theta = 10.6$ | 0 |
| | | | | | | | | Gumbel | $\theta = 3.58$ | 0 |
| | | | | | | | | Plackett | $\theta = 39.04$ | 0 |
| | | | | | | | | Student's $t$ | $\vartheta = 8,\ \rho_{s,d} = 0.98$ | 0.088 |

Note: Higher $p_{value}$ indicates better fit.

Table S5. Field significance tests for the changes in drought severity corresponding to 10-year return period

| Trends | Region | D1-Out 1970-2013 vs. 1926-69 | | | | D1-No Out 1980-2009 vs. 1950-1979 | | | | D2-No Out 1980-2009 vs. 1950-1979 | | | |
|---|---|---|---|---|---|---|---|---|---|---|---|---|---|
| | | Upward | | Downward | | Upward | | Downward | | Upward | | Downward | |
| | | $p_{fdr}$ | $N$ | $p_{fdr}$ | $N$ | $p_{fdr}$ | $N$ | $p_{fdr}$ | $N$ | $p_{fdr}$ | $N$ | $p_{fdr}$ | $N$ |
| Median | Central | 0.012 | 14 | 0.001 | 5 | 4.88e$^{-4}$ | 0 | 4.88e$^{-4}$ | 4 | 0.0014 | 5 | 0.021 | 32 |
| | ENC | 0.027 | 4 | 0.042 | 6 | 4.88e$^{-4}$ | 0 | 9.7e$^{-4}$ | 4 | 9.76e$^{-4}$ | 3 | 0.021 | 21 |
| | Northeast | 0.009 | 12 | 4.88e$^{-4}$ | 0 | 0.027 | 9 | 0.003 | 2 | 0.016 | 3 | 0.021 | 18 |
| | Northwest | 0.016 | 3 | 4.88e$^{-4}$ | 0 | 4.88e$^{-4}$ | 0 | 0.002 | 2 | 4.88e$^{-4}$ | 0 | 0.012 | 16 |
| | Southeast | 0.001 | 4 | 4.88e$^{-4}$ | 4 | 0.021 | 0 | 9.7e$^{-4}$ | 2 | 4.88e$^{-4}$ | 0 | 0.027 | 15 |
| | South | 0.016 | 15 | 0.009 | 8 | 4.88e$^{-4}$ | 0 | 0.012 | 11 | 0.0092 | 2 | 0.021 | 27 |
| | Southwest | 0.012 | 5 | 4.88e$^{-4}$ | 7 | 9.76e$^{-4}$ | 1 | 4.88e$^{-4}$ | 0 | 0.0024 | 2 | 0.009 | 10 |
| | West | 0.003 | 1 | - | - | - | - | 0.016 | 1 | 0.003 | 0 | 0.012 | 8 |
| | WNC | 0.001 | 10 | 0.002 | 11 | 0 | 0 | 0.034 | 12 | 4.88e$^{-4}$ | 1 | 0.021 | 32 |
| Slope | Central | 0.0003 | 13 | 0.006 | 7 | 2.85e$^{-5}$ | 5 | 0.031 | 3 | 0.0144 | 6 | 0.031 | 27 |
| | ENC | 0.040 | 3 | 0.003 | 8 | 1.19e$^{-8}$ | 1 | 0.006 | 4 | 0.009 | 15 | 0.008 | 15 |
| | Northeast | 2.61e$^{-6}$ | 6 | 0.002 | 1 | 0.001 | 7 | 0.004 | 0 | 0.0087 | 4 | 0.016 | 13 |
| | Northwest | 0.022 | 5 | 0.003 | 4 | 3.44e$^{-8}$ | 0 | 6.12e$^{-10}$ | 1 | 7.7e$^{-4}$ | 2 | 0.027 | 14 |
| | Southeast | 0.0006 | 5 | 0.021 | 5 | 7.64e$^{-5}$ | 1 | 0.027 | 4 | 0.026 | 5 | 0.039 | 13 |
| | South | 0.005 | 17 | 0.035 | 9 | 0.019 | 6 | 0.011 | 5 | 0.029 | 16 | 0.032 | 19 |
| | Southwest | 0.004 | 6 | 0.044 | 7 | 0.025 | 5 | 0.0012 | 0 | 0.037 | 11 | 0.002 | 9 |
| | West | 6.45e$^{-10}$ | 0 | - | - | - | - | 5.19e$^{-4}$ | 1 | 2.9e$^{-4}$ | 0 | 0.006 | 6 |
| | WNC | 0.011 | 7 | 0.002 | 10 | 0.022 | 1 | 0.009 | 8 | 0.025 | 10 | 0.009 | 26 |

**Table S6.** Field significance tests for the changes in drought severity corresponding to 100-year return period

| Trends | Region | D1-Out 1970-2013 vs. 1926-69 | | | | D1-No Out 1980-2009 vs. 1950-1979 | | | | D2-No Out 1980-2009 vs. 1950-1979 | | | |
| | | Upward | | Downward | | Upward | | Downward | | Upward | | Downward | |
| | | $p_{fdr}$ | $N$ | $p_{fdr}$ | $N$ | $p_{fdr}$ | $N$ | $p_{fdr}$ | $N$ | $p_{fdr}$ | $N$ | $p_{fdr}$ | $N$ |
| --- | --- | --- | --- | --- | --- | --- | --- | --- | --- | --- | --- | --- | --- |
| Median | Central | 0.021 | 15 | 0.021 | 4 | 0.042 | 5 | 0.001 | 5 | 0.021 | 7 | 0.003 | 25 |
| | ENC | 0.021 | 5 | 0.042 | 6 | $4.88e^{-4}$ | 0 | $4.88e^{-4}$ | 0 | 0.016 | 3 | 0.012 | 17 |
| | Northeast | 0.027 | 14 | 0.002 | 1 | $4.88e^{-4}$ | 0 | 0.002 | 4 | 0.003 | 3 | 0.042 | 18 |
| | Northwest | 0.027 | 4 | $4.88e^{-4}$ | 0 | $4.88e^{-4}$ | 0 | $9.77e^{-4}$ | 2 | $4.88e^{-4}$ | 0 | 0.012 | 15 |
| | Southeast | 0.021 | 5 | 0.002 | 3 | - | - | 0.003 | 3 | $4.88e^{-4}$ | 0 | 0.034 | 15 |
| | South | 0.007 | 18 | $4.88e^{-4}$ | 7 | $4.88e^{-4}$ | 0 | 0.012 | 12 | 0.021 | 6 | 0.027 | 25 |
| | Southwest | 0.042 | 8 | 0.012 | 7 | $9.76e^{-4}$ | 1 | 0.009 | 1 | 0.027 | 4 | 0.016 | 11 |
| | West | 0.003 | 3 | - | - | - | - | 0.001 | 1 | $4.88e^{-4}$ | 0 | 0.016 | 8 |
| | WNC | 0.027 | 11 | 0.012 | 11 | - | - | 0.034 | 14 | 0.009 | 4 | 0.016 | 28 |
| Slope | Central | 0.018 | 13 | $5.97e^{-4}$ | 7 | $1.42e^{-5}$ | 5 | $6.13e^{-5}$ | 3 | 0.011 | 16 | 0.043 | 24 |
| | ENC | 0.003 | 3 | $7.63e^{-4}$ | 8 | $2.54e^{-9}$ | 1 | $1.03e^{-5}$ | 4 | 0.003 | 19 | 0.009 | 13 |
| | Northeast | 0.011 | 8 | 0.0056 | 2 | $7.5e^{-4}$ | 7 | 0.032 | 3 | 0.008 | 7 | $5.15e^{-5}$ | 12 |
| | Northwest | 0.007 | 5 | 0.002 | 4 | $3.41e^{-8}$ | 0 | $3.9e^{-12}$ | 1 | 0.005 | 5 | 0.005 | 10 |
| | Southeast | 0.002 | 5 | 0.002 | 5 | 0.030 | 2 | $9.19e^{-4}$ | 4 | 0.040 | 12 | 0.019 | 8 |
| | South | 0.018 | 19 | 0.010 | 8 | 0.031 | 5 | 0.041 | 10 | 0.025 | 22 | 0.001 | 17 |
| | Southwest | 0.017 | 7 | 0.032 | 7 | 0.034 | 4 | $1.04e^{-4}$ | 1 | 0.037 | 11 | 0.002 | 9 |
| | West | 0.003 | 1 | - | - | - | - | $2.13e^{-5}$ | 1 | 0.009 | 2 | 0.007 | 5 |
| | WNC | 0.018 | 10 | $7.67e^{-4}$ | 10 | - | - | 0.031 | 10 | 0.013 | 14 | 0.004 | 23 |

**Table S7.** Field significance tests for the changes in variance in drought severity between 1970-2013 and 1926-69

| Region | Upward | | Downward | |
|---|---|---|---|---|
| | $p_{fdr}$ | $N$ | $p_{fdr}$ | $N$ |
| *Central* | *0.0024* | *1* | *0.0068* | *9* |
| ENC | 0.0013 | 1 | 0.0026 | 1 |
| *Northeast* | *0* | *0* | *0.011* | *9* |
| Northwest | $3.16e^{-4}$ | 0 | 0.0042 | 4 |
| Southeast | 0.0034 | 3 | 0.0056 | 3 |
| South | $1.73e^{-6}$ | 1 | $9.5e^{-4}$ | 0 |
| Southwest | 0 | 0 | $5.78e^{-5}$ | 0 |
| West | 0.0095 | 2 | 0.0045 | 1 |
| WNC | 0.0061 | 7 | $7.86e^{-4}$ | 1 |

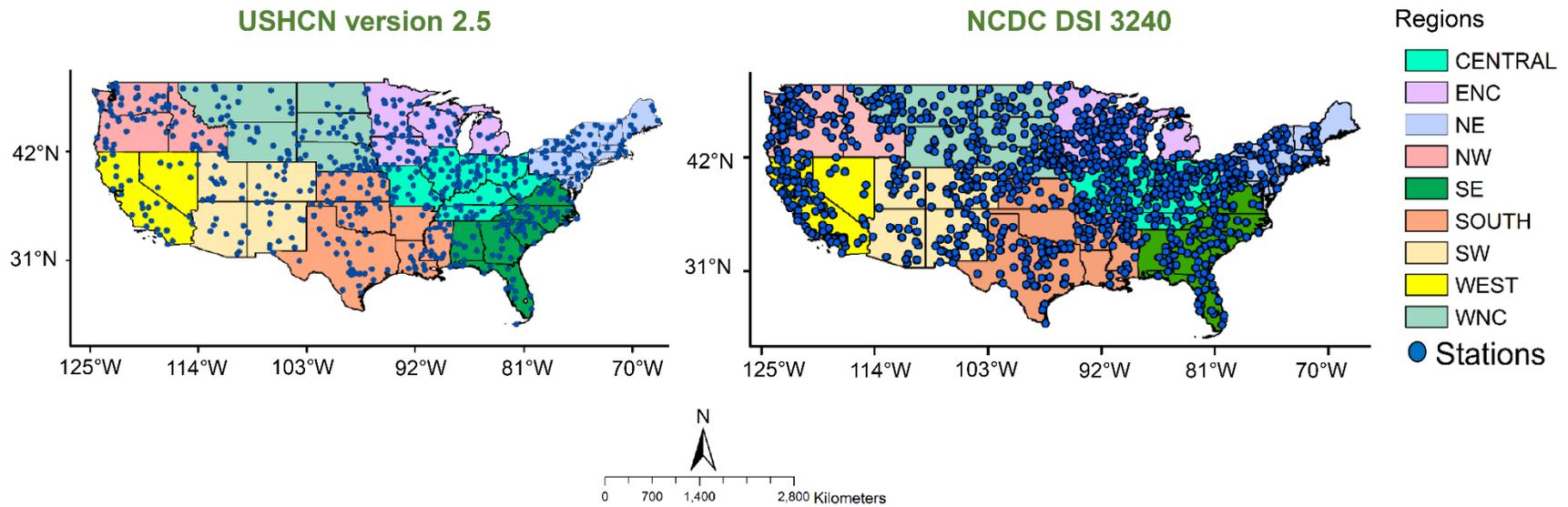

**Fig. S1.** Spatial distributions of meteorological stations over nine NCDC climate regions in USHCN version 2.5 (*left*) and NCDC DSI-3240 (*right*) datasets. Nine regions are shown with nine different colors.

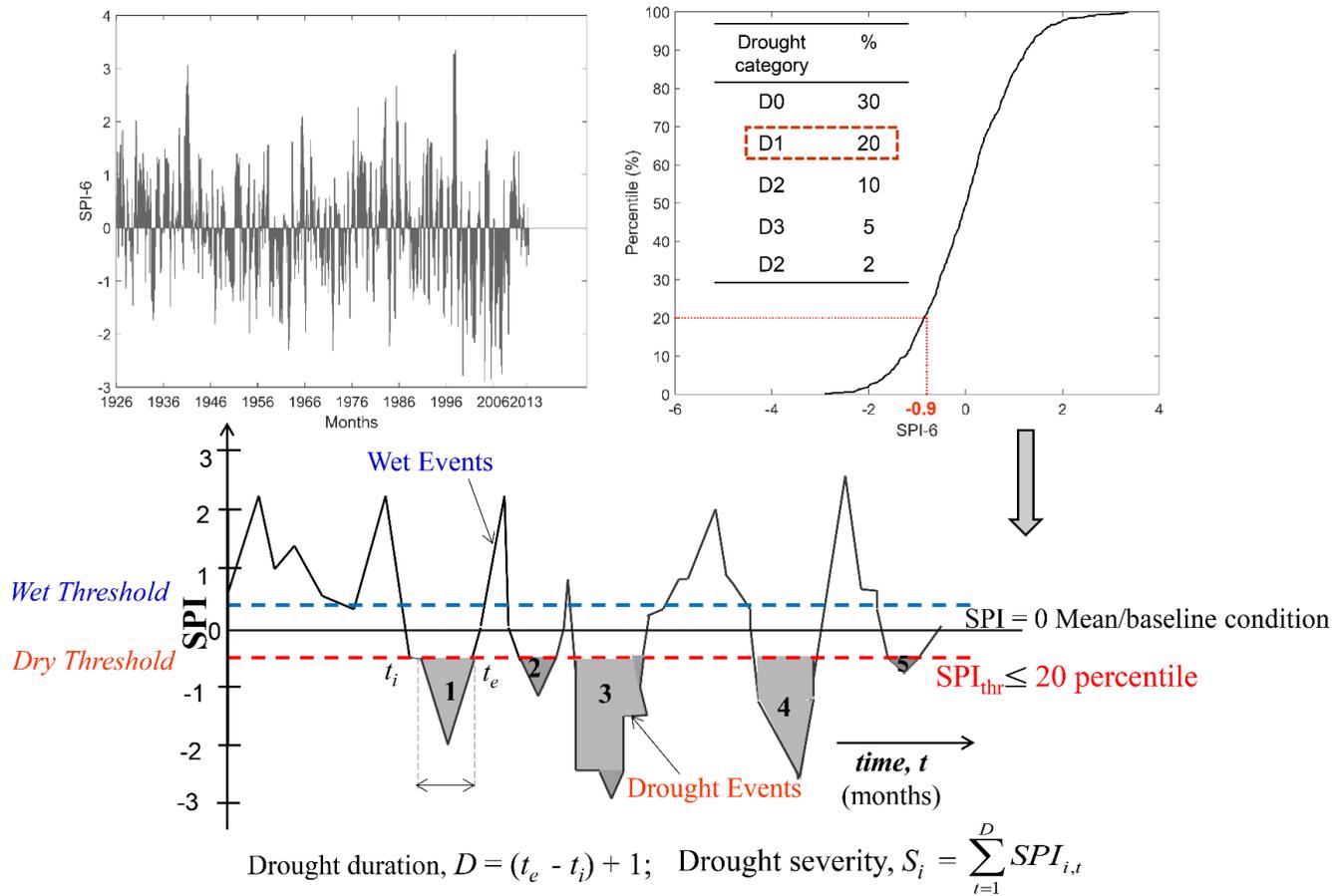

$$\text{Drought duration, } D = (t_e - t_i) + 1; \quad \text{Drought severity, } S_i = \sum_{t=1}^{D} SPI_{i,t}$$

**Fig. S2.** Identification of drought events and associated properties using threshold method from SPI time series. For illustration purpose, a location (longitude -117.08º and latitude 32.64º) in California is chosen. The 20[th] percentile threshold provides SPI value of -0.9 (*top panel*, *left*) during 1926 – 2013 for this station. Drought events are shown in shaded color (*bottom panel*). Drought classes are defined (Svoboda, 2002) based on SPI percentiles (*D*2: 5–≤ 10 percentile, *D*3: 2–≤ 5 percentile, and *D*4: ≤ 2 percentile as used by the US drought monitor).

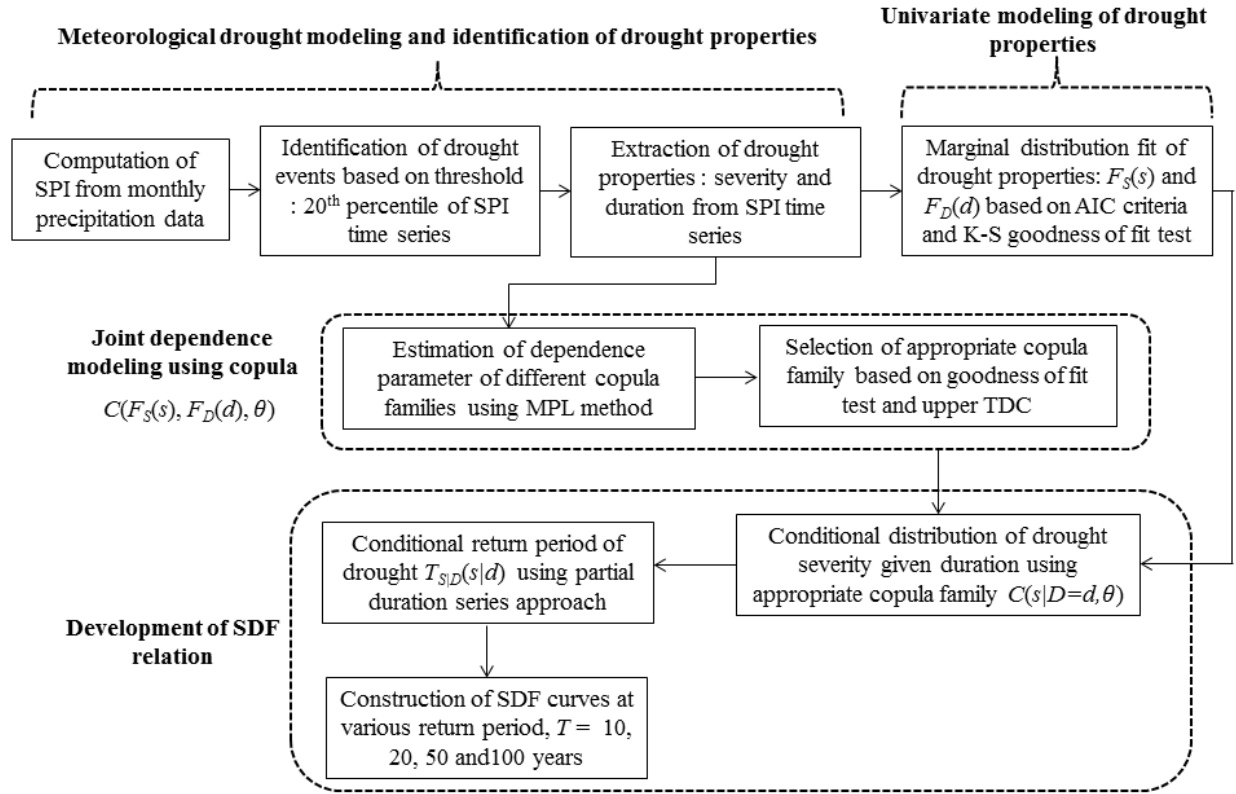

**Fig. S3.** Overview of the construction of SDF curves using copula-based conditional simulation.

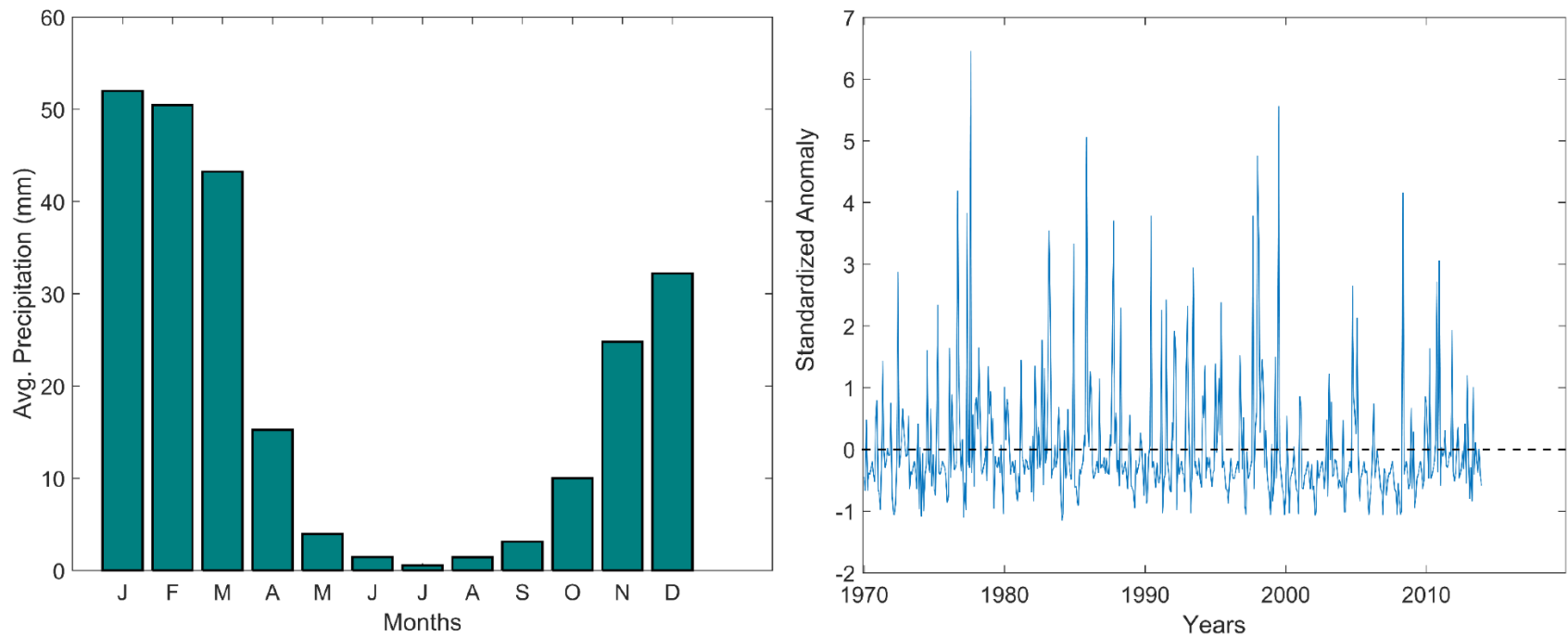

**Fig. S4.** Standardized anomaly in monthly precipitation data in a station location (longitude -117.08° and latitude 32.64°) in California.

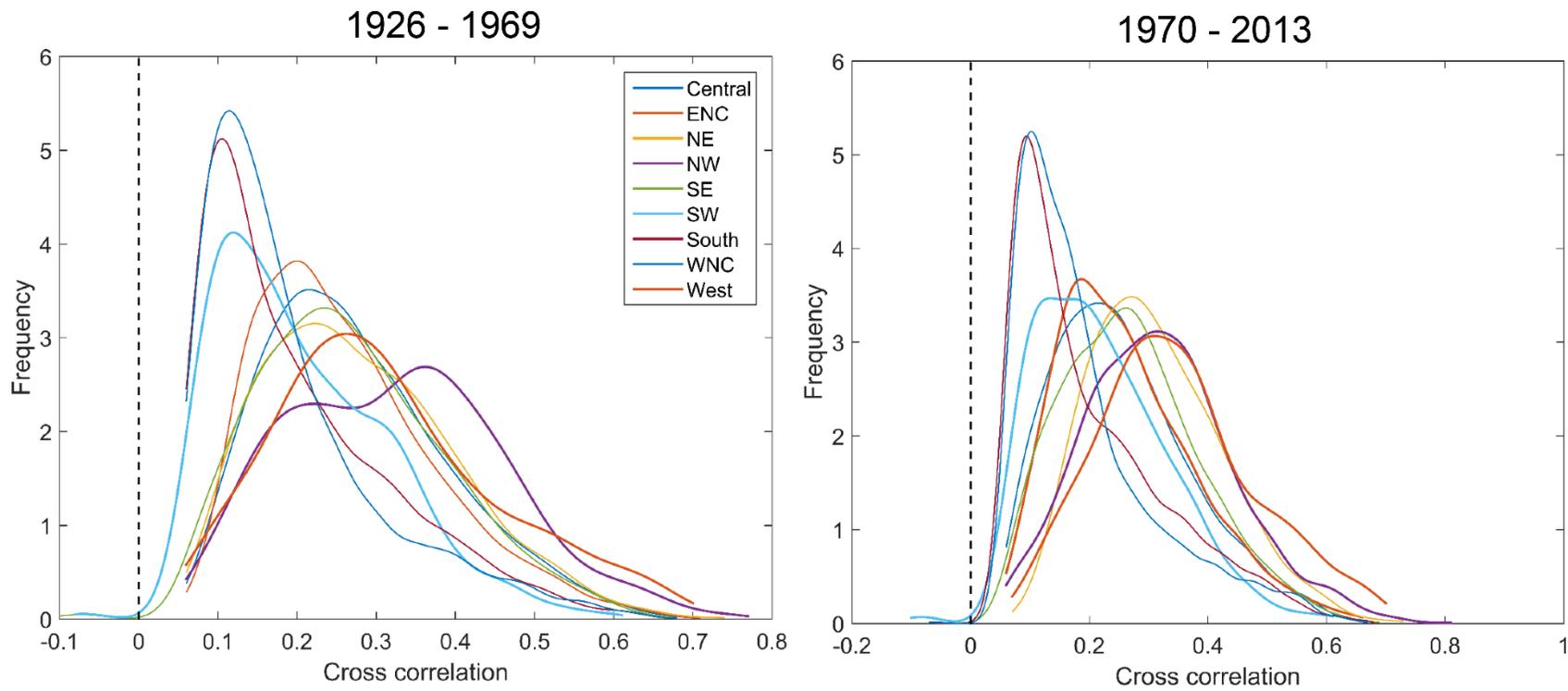

**Fig. S5.** Distributions of (significant) Kendall's τ cross-correlation for the two 44-year time periods in D1- Out data. Statistical significance is computed at α = 0.05 significance level. The legend applies to both panels.

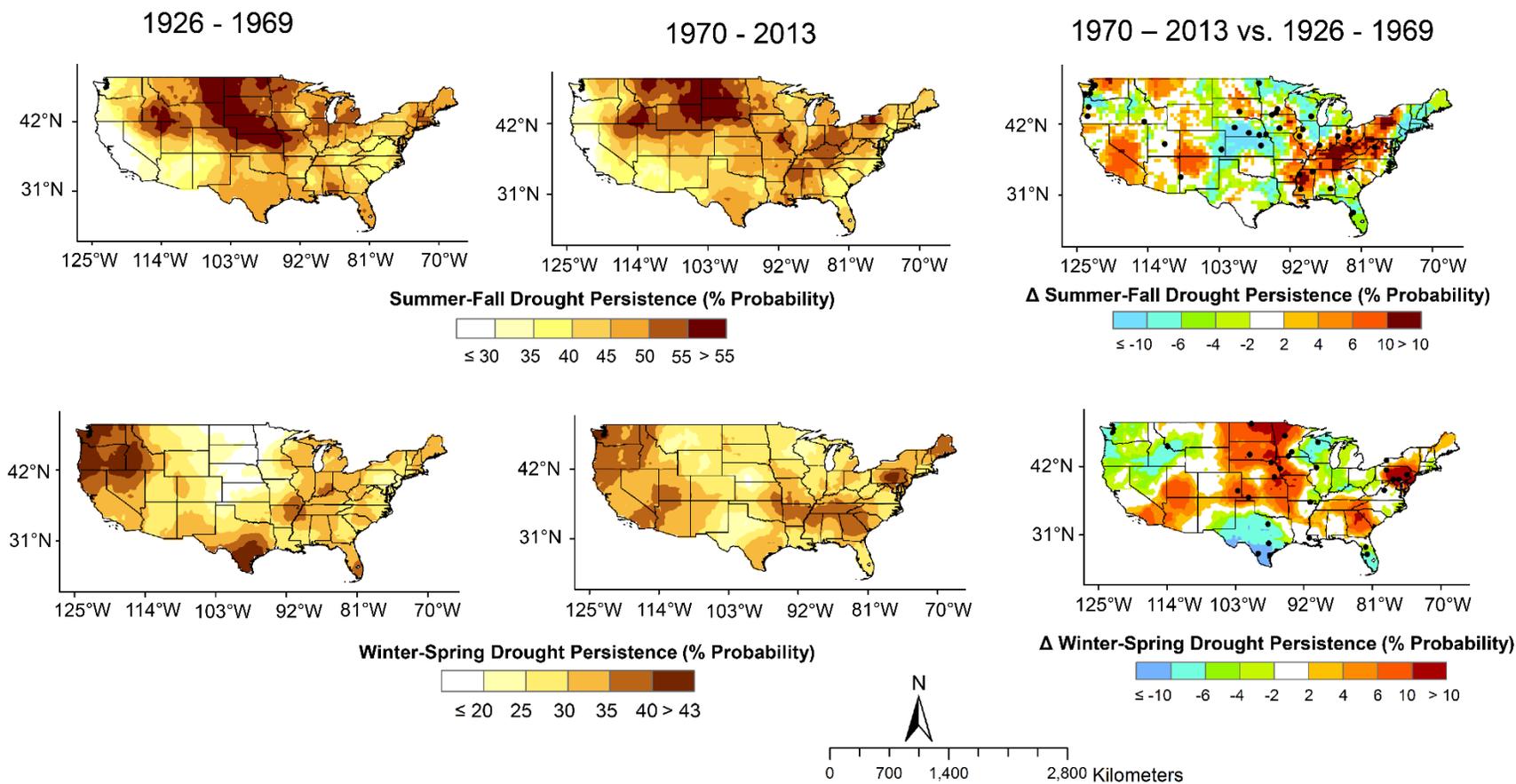

**Fig. S6.** Spatial distributions of seasonal persistence probability in two consecutive time-windows 1926-1969 (left), 1970-2013 (middle) and the corresponding difference map (right) comparing 1970-2013 versus 1926-1969 for summer to fall (*top*) and winter to spring (*bottom*) droughts. The black filled circle in the plot indicates locally significant differences in the persistence probability.

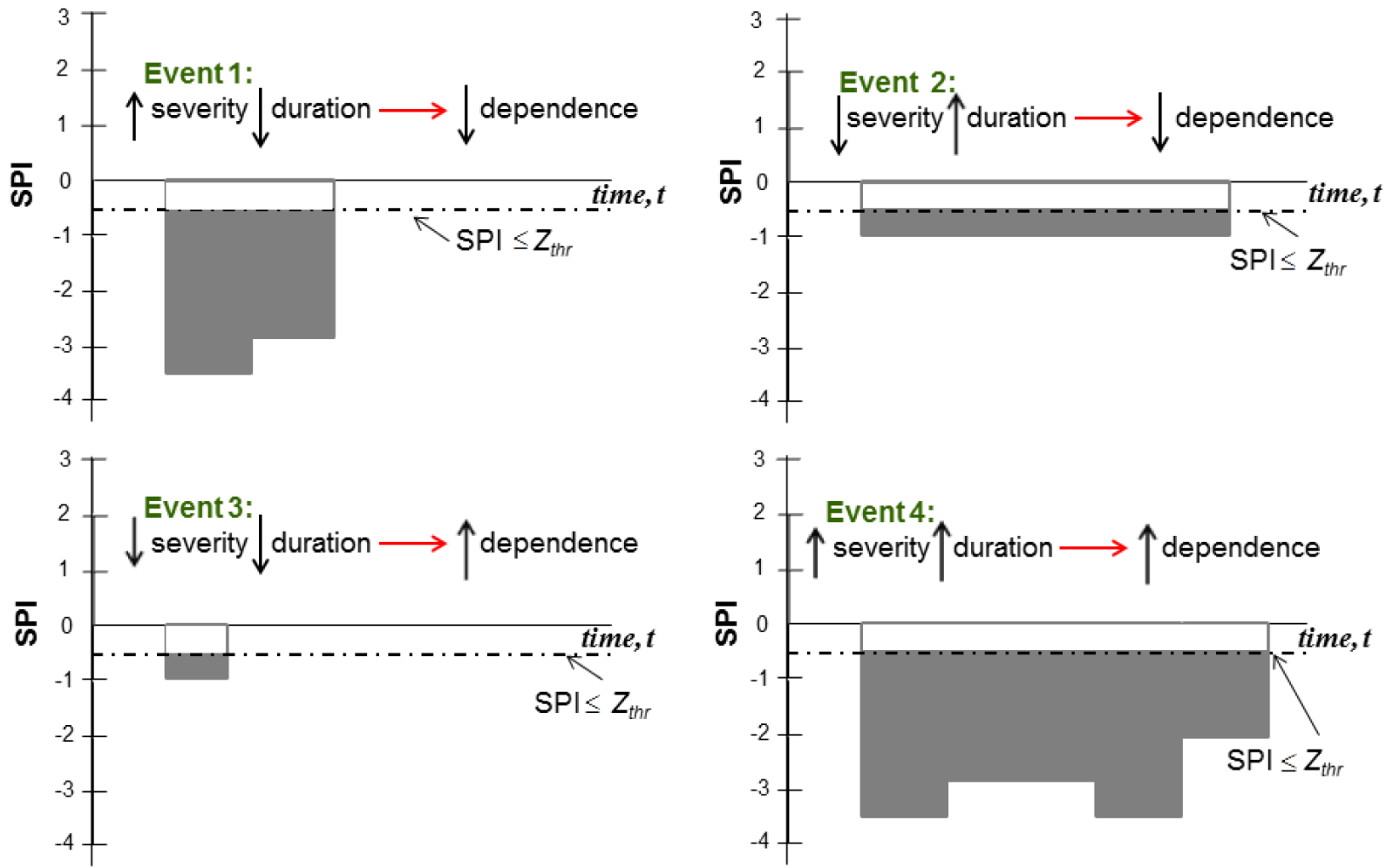

**Fig. S7.** Hypothetical framework to assess dependence pattern that may emerge from associated drought properties.

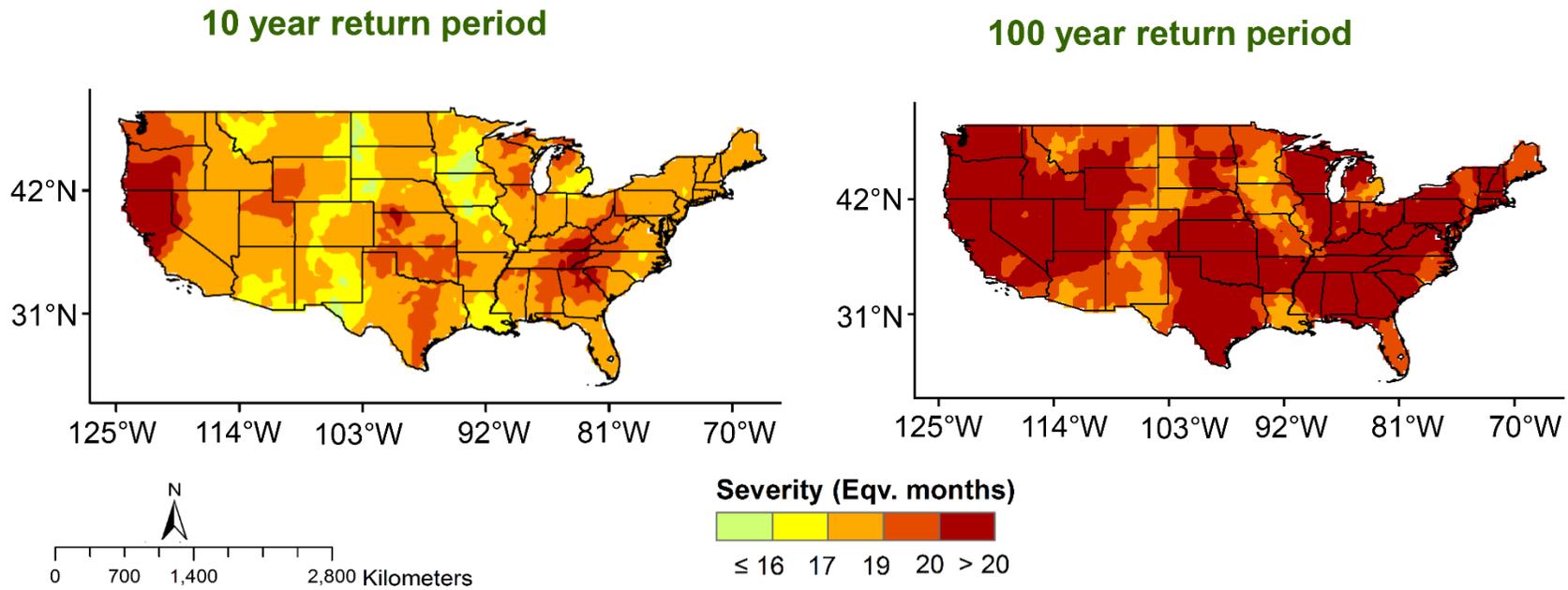

**Fig. S8.** Spatial distributions of drought severity for drought durations of 12-months. 10- (*left*), and 100- (*right*) year return periods are obtained using SDF relations.

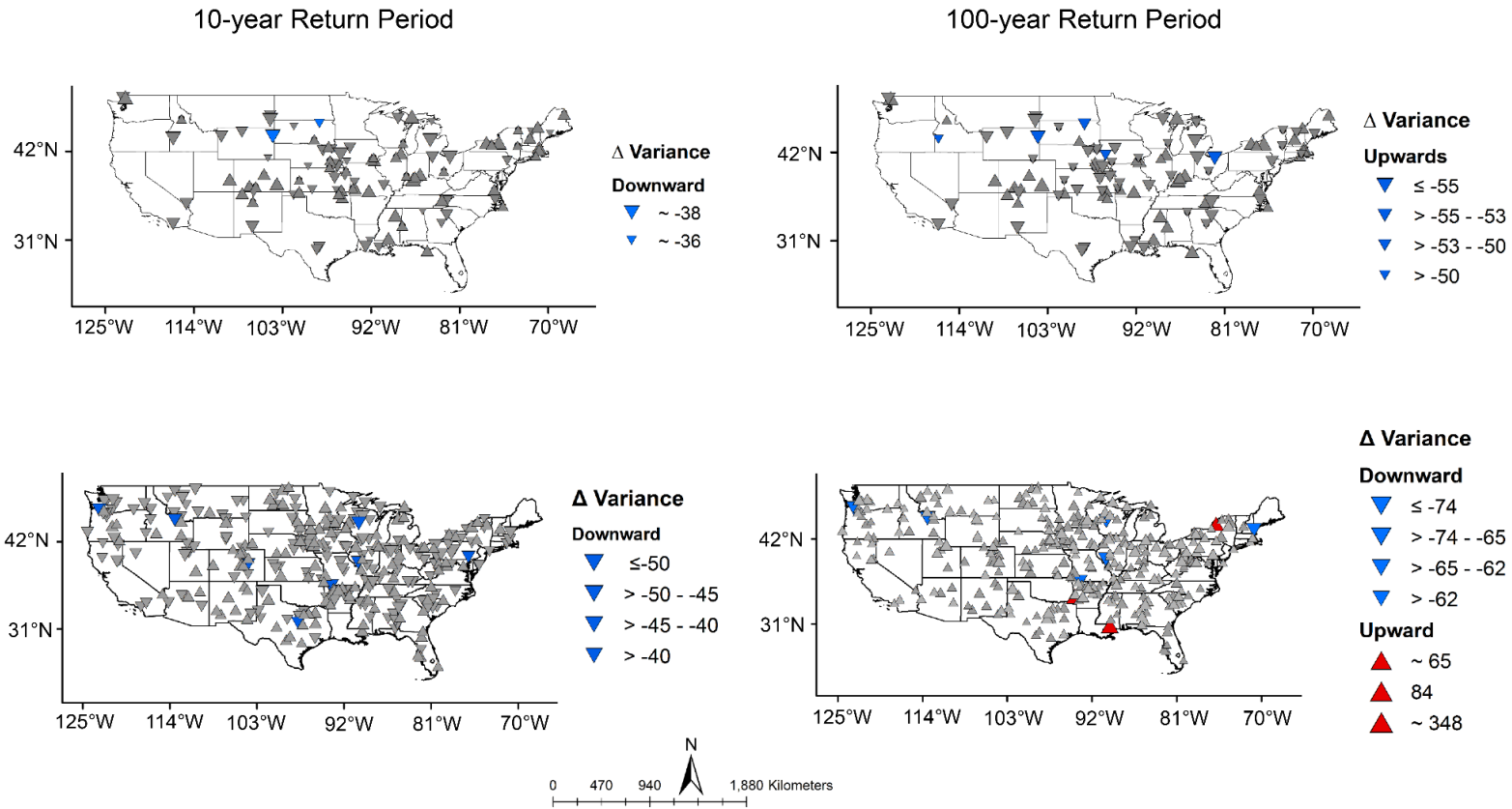

**Fig. S9.** Changes (1980 – 2009 versus 1950 – 1979) in the (temporal) variance of drought severity corresponding to 10- (*left*) and 100- (*right*) year return periods in D1- No Out (*top*) and D2- No Out (*bottom*) datasets. Triangles indicate geographical locations of stations and colors describe the sign and the significance of the estimated trend. The upward, downward and insignificant trends are marked with red, blue and gray colored triangles. The size (and shading) of the triangle is proportional to the magnitude of the trend.